\documentclass[a4paper,11pt]{article}
\pdfoutput=1
\usepackage{jheppub}
\pdfoutput=1

\usepackage{epsf}
\usepackage{epsfig}
\usepackage{subfig}
\usepackage{mathtools}
\usepackage{hhline}
\usepackage{float}
\usepackage{multirow}
\usepackage{nicefrac}
\usepackage{epstopdf}
\usepackage{slashed}
\usepackage{xcolor}
\usepackage{url}  
\usepackage{breqn}
\usepackage{amsmath}
\usepackage{mathtools}
\usepackage{graphicx}
\usepackage{feynmp-auto}

\usepackage{braket}
\usepackage{subdepth}
\graphicspath{ {figures/} }

\newcommand{\be}{\begin{eqnarray}}
\newcommand{\ee}{\end{eqnarray}}

\newcommand{\ba}{\begin{array}}
\newcommand{\ea}{\end{array}}
\newcommand{\bee}{\begin{equation}\ba{c}}
\newcommand{\eee}{\ea\end{equation}}

\newcommand{\bi}{\begin{itemize}}
\newcommand{\ei}{\end{itemize}}

\newcommand{\ffrac}[2]{\ensuremath{\frac{\displaystyle #1}{\displaystyle #2}}}

\allowdisplaybreaks

\def\Li{\text{Li}}

\def\be{\begin{equation}}
\def\ee{\end{equation}}
\def\bea{\begin{eqnarray}}
\def\eea{\end{eqnarray}}

\def\bbuildrel#1_#2^#3{\mathrel{\mathop{\kern 0pt#1}\limits_{#2}^{#3}}}
\def\slash#1{\setbox0=\hbox{$#1$}#1\hskip-\wd0\dimen0=5pt\advance
       \dimen0 by-\ht0\advance\dimen0 by\dp0\lower0.5\dimen0\hbox
         to\wd0{\hss\sl/\/\hss}}

\def\be{\begin{equation}}
\def\ee{\end{equation}}
\def\beq{\begin{eqnarray}}
\def\eeq{\end{eqnarray}}

\def\eps{\epsilon}

\def\slash#1{#1 \hskip-0.45em /}

\def\DB0{\partial B_0}

\newcommand{\lsim}
{\;\raisebox{-.3em}{$\stackrel{\displaystyle <}{\sim}$}\;}
\newcommand{\gsim}
{\;\raisebox{-.3em}{$\stackrel{\displaystyle >}{\sim}$}\;}

\def\Cl2{\mbox{Cl}_2}

\def\slash#1{#1 \hskip-0.45em /}

\newcommand{\hs}[1]{\hspace*{#1 pt}}

\newcommand{\MB}[2]{\hs{-12} \int\limits_{\hs{15}_{ #1 -i \,
\infty}}^{\hs{15}^{ #1 +i\, \infty}} \hs{-15} \frac{d #2}{2\pi i}}

\usepackage{booktabs}
\usepackage{siunitx}

\usepackage{color}

\definecolor{Brown}{rgb}{0.5,0.25,0}

\title{Inclusive $\bar{B} \to X_s \ell^+ \ell^-$ with a hadronic mass cut}
\author{Tobias Huber$^1$,}
\author{Tobias Hurth$^2$,}
\author{Jack Jenkins$^3$,}
\author{Enrico Lunghi$^3$}
\affiliation{
$^1$Theoretische Physik 1, Center for Particle Physics Siegen (CPPS), Universit\"at Siegen, \\ Walter-Flex-Stra{\ss}e 3, D-57068 Siegen, Germany\\
$^2$PRISMA+ Cluster of Excellence and Institute for Physics (THEP), \\ Johannes Gutenberg University, D-55099 Mainz, Germany\\
$^3$Physics Department, Indiana University, Bloomington, IN 47405, USA
}

\emailAdd{huber@physik.uni-siegen.de}
\emailAdd{tobias.hurth@cern.ch}
\emailAdd{jackjenk@iu.edu}
\emailAdd{elunghi@indiana.edu}

\abstract{
The hadronic mass spectrum of inclusive $\bar{B} \to X_s \ell^+ \ell^-$ is investigated at next to leading order in the heavy quark expansion. For mild cuts on the hadronic mass, the expansion, which applies when the cut is released, remains convergent. However, the cuts used at BaBar and Belle to reduce backgrounds from charged current semileptonic processes are too severe for a description in terms of matrix elements of local operators to apply. Strategies for interpolating between the two regions are discussed.
}

\preprint{
\begin{minipage}{3cm}
\small
\flushright
SI-HEP-2023-11\\
P3H-23-037\\
MITP/23-022
\end{minipage}}

\begin{document}

\maketitle


\section{Introduction}
\label{sec:introduction}

The inclusive rare decay $\bar{B} \to X_s \ell^+ \ell^-$ $(\ell = e, \mu)$ proceeds through a flavor changing neutral current (FCNC), which is forbidden in the Born approximation of the Standard Model (SM). However, the underlying $b \to s$ transition occurs at higher order through simultaneous emission of a pair of charged gauge bosons, or emission of a neutral gauge boson from a virtual charged gauge boson or quark. Therefore $\bar{B} \to X_s \ell^+ \ell^-$ directly probes the quantum fluctuations of the SM at the electroweak scale, and is sensitive to potential physics beyond the SM~\cite{Hurth:2010tk, Hurth:2003vb}. 

Tensions between measurements and SM predictions in precision B physics, known as the B anomalies, have persisted over the last decade, mainly driven by LHCb results on branching fractions and angular observables of exclusive modes with muons such as $\bar{B} \to K \mu^+ \mu^-$ and $\bar{B} \to K^* \mu^+ \mu^-$~\cite{LHCb:2020gog, LHCb:2020lmf, LHCb:2016eyu, LHCb:2016ykl,LHCb:2015svh, LHCb:2021xxq, LHCb:2021zwz,LHCb:2015wdu, LHCb:2018jna, LHCb:2015tgy}. Such observables suffer from power corrections which are difficult to access, among them hadronic contributions from virtual charm quarks which are not described within QCD factorization. While these nonlocal matrix elements can be addressed with analyticity methods~\cite{Gubernari:2022hxn}, it is currently not possible to confidently separate new physics (NP) from them; the NP significance of the tensions presently depends on order of magnitude estimates for these unknown long distance effects~\cite{Hurth:2022lnw,Hurth:2020ehu}.

On the other hand, the ratios $R_K$ and $R_{K^*}$ of branching fractions into muons compared to electrons are theoretically very clean, with uncertainties of less than one percent and central values close to unity in the SM due to lepton flavor universality~\cite{Hiller:2003js, Bordone:2016gaq}. Tensions in $R_K$ and $R_{K^*}$ were reported at the level of $3\sigma$ by LHCb~\cite{LHCb:2017avl,LHCb:2021trn}. The crucial issue in this context is that the tensions were rather consistent with the previously found tensions in the angular observables; the persistence of the tensions in $R_K$ and $R_{K^*}$ supported the NP interpretation in the other exclusive observables, despite their sensitivity to long distance physics~\cite{Hurth:2021nsi}. However, the anomaly has evaporated in the latest LHCb measurement~\cite{LHCb:2022zom} which is now one of the most precise measurements in FCNC transitions, and along with it, the option to disentangle NP in the exclusive modes with a clean observable.

Since $R_{K}$ and $R_{K^*}$ are now consistent with the SM, investigating the inclusive mode $\bar{B} \to X_s \mu^+ \mu^-$ is the only remaining option to resolve the persisting anomalies in the exclusive modes such as $\bar{B} \to K \mu^+ \mu^-$ and $\bar{B} \to K^* \mu^+ \mu^-$. Inclusive and exclusive decays offer complementary information in the search for NP in $b \to s$ transitions~\cite{Huber:2020vup}. However, inclusive decays are much cleaner, as they are analyzed in the operator product expansion (OPE) in terms of a handful of local matrix elements, and bounds on nonlocal power corrections originating from resolved virtual photons can be calculated within Soft Collinear Effective Theory (SCET)~\cite{Hurth:2017xzf,Benzke:2017woq,Benzke:2020htm}. 

In the future the inclusive mode will be measured at Belle II with high precision~\cite{Belle-II:2018jsg}. A semi-inclusive measurement might  be possible even at LHCb~\cite{Amhis:2021oik}.
Presently, inclusive measurements are available from BaBar~\cite{BaBar:2004mjt,BaBar:2013qry} and Belle~\cite{Belle:2005fli,Belle:2014owz} with combined statistics on electrons and muons. The results\footnote{$[q_1^2, q_2^2]$ indicates the bin $q_1^2 < m_{\ell \ell}^2 < q_2^2$ in units of ${\rm GeV}^2$.}
\begin{align}
\mathcal{B}[1,6]_{\ell \ell}^{\rm exp} = \left\{ \begin{array}{ll}
(1.49\pm0.50{}^{+0.41}_{-0.32}) \times 10^{-6} \indent & {\rm Belle}
\vspace{0.3cm}\\
(1.60\,{}^{+0.41}_{-0.39}{}^{+0.17}_{-0.13}\pm 0.18) \times 10^{-6} \indent & {\rm BaBar}
\end{array} \right.
\label{eq:BaBar}
\end{align}
are compatible with each other and with the SM predictions~\cite{Huber:2020vup}
\begin{align}
  \mathcal{B}[1,6]_{ee}^{\rm SM} = (1.78\pm0.13)\times10^{-6}\, , \indent \mathcal{B}[1,6]_{\mu\mu}^{\rm SM} = (1.73\pm0.13)\times10^{-6} \, .
  \label{eq:bsllSM}
\end{align}
The first two uncertainties indicated in eq.~\eqref{eq:BaBar} are statistical and systematic, respectively. The final uncertainty on the BaBar result accounts, separately from other systematics, the reconstruction of missing modes using a sum-over-exclusive tagging method. The missing modes include modes removed by a cut $M_X < 1.8\,{\rm GeV}$ (for Belle $M_X < 2.0 \, {\rm GeV}$) necessary to reduce double semileptonic backgrounds from same and/or opposite side decays ${\bar{B} \to X_c( \to X \ell^+ \nu) \ell^- \nu}$ and $\Bar{B}(\to X_c \ell^- \nu)B (\to X_{\bar{c}} \ell^+ \nu)$ respectively. With sufficient statistics at Belle II, it may be feasible to use the recoil tagging method, in which the kinematics of the $X_s$ system is determined indirectly by tagging the fully hadronic decays of the partner $B$ meson and the lepton momenta in $\bar{B} \to X \ell^+ \ell^-$. However, a cut on the hadronic mass will probably still be necessary to reduce backgrounds from ${\bar{B} \to X_c( \to X \ell^+ \nu) \ell^- \nu}$ on the signal side.

At BaBar and Belle, the effect of the cut on the hadronic invariant mass was taken into account with a signal Monte Carlo formed by smearing the spectrum of $b\to s (g) \ell^+ \ell^-$ with a Gaussian Fermi motion model~\cite{Ali:1998nq}. Alternatively, measurements of $\bar{B} \to X_s \ell^+ \ell^-$ with a hadronic mass cut can be compared to the corresponding theory predictions, using the framework of SCET to address the multi-scale problem introduced by the hadronic mass cut. At leading order in $1/m_b$, there is a single shape function which is universal to all heavy-to-light-current B decays~\cite{Neubert:1993um, Bernlochner:2020jlt}. These shape functions represent the soft functions in the factorization within SCET and are well-defined HQET (Heavy Quark Effective Theory) matrix elements. The effect of the hadronic mass cut was first analyzed in~\cite{Lee:2005pwa}, but with some simplifications and certain problems about the SCET scaling of the virtual photon in the low dilepton-mass region as indicated in~\cite{Bell:2010mg,Hurth:2017xzf,Benzke:2017woq}. At order $1/m_b$, five subleading shape functions appear, and enter with different kernels in SCET convolution integrals for various heavy-to-light decays. The uncertainty due to these subleading shape functions is presently estimated at $5-10\%$~\cite{Lee:2005pwa,Lee:2008xc}. It might be possible to reduce this uncertainty by incorporating more information on moments of the subleading shape functions within HQET (see e.g.~\cite{Gunawardana:2019gep}).

In this article we follow another strategy to reduce the uncertainty due to the hadronic mass cut. We consider the effect of the hadronic mass cut for {mild}  cuts in the OPE region and  analyze  the validity of the OPE by studying the explosion of power corrections as the cut is lowered into the shape function region (${M_X^{\rm cut} \sim \sqrt{\Lambda m_b}}$).  Since the OPE region does not overlap with the cuts in the shape function region required by experiment, in the future, as a next step an interpolation between the OPE region and the shape function region using SCET is planned. Ratios of observables with the same hadronic mass cut are suitable for investigating the interpolation. Certain $\bar{B} \to X_s \ell^+ \ell^-$ observables are already known to be independent of the shape function in SCET at least at leading order in $1/m_b$, such as the zero-crossing of the forward-backward asymmetry~\cite{Bell:2010mg}.
 
For this purpose, we compute the fully differential distribution of $\bar{B} \to X_s \ell^+ \ell^-$ at $O(\alpha_s)$ in the OPE. We also compute those power corrections at $O(\alpha_s/m_b^2)$ which are the most divergent in $1/M_X^{\rm cut}$ and use them as an indicator for the breakdown of the OPE. Moreover, the three $\bar{B} \to X_s \ell^+ \ell^-$ angular observables, together with the $\bar{B} \to X_u \ell^- \nu$ branching fraction, all with the same hadronic mass cut, constitute a basis of four heavy-to-light-current observables from which three normalized observables, which are both sensitive to NP and rather independent of the hadronic mass cut, can be constructed. We anticipate that both perturbative and nonperturbative corrections are essentially eliminated in these  ratios in the OPE region.

The organization of this paper is as follows. In section~\ref{sec:theo}, the effective Lagrangian and angular decomposition of $\bar{B} \to X_s \ell^+ \ell^-$ is reviewed. In section~\ref{sec:numerics}, results for the effect of the hadronic mass cut on the rate and angular observables of $\bar{B} \to X_s \ell^+ \ell^-$ are presented. We summarize in section~\ref{sec:summary} and relegate technical details to appendices~\ref{sec:analyticresults} and~\ref{sec:plus}.


\section{Theoretical framework}
\label{sec:theo}

Integrating out electroweak gauge bosons, the top quark and Higgs boson from the SM leads to an effective Lagrangian
\begin{align}
\mathcal{L}(b \to s \ell^+ \ell^-) &= \frac{4 G_F}{\sqrt{2}} V_{ts}^* V_{tb} \sum_{i=1}^{10} C_i Q_i \, ,
\label{eq:Lagrangian}
\end{align}
where $Q_{3\dots 6}$ are QCD penguin operators and
\begin{align}
Q_1 &= (\bar{s} \gamma_\mu P_L T^a c)(\bar{c} \gamma^\mu P_L T^a b) \, , & Q_2 &= (\bar{s} \gamma_\mu P_L c)(\bar{c} \gamma^\mu P_L  b)\, , \\
Q_7 &= \frac{e m_b}{16\pi^2} (\bar{s} \sigma^{\mu \nu} P_R b) F_{\mu \nu} \, , & Q_8 &= \frac{g m_b}{16\pi^2} (\bar{s} \sigma^{\mu \nu} P_R T^a b) G^a_{\mu \nu} \, , \nonumber \\
Q_9 &= \frac{\alpha}{4\pi} (\bar{s}\gamma_\mu P_L b) (\bar{\ell} \gamma^\mu \ell) \, , & Q_{10} &= \frac{\alpha}{4\pi}(\bar{s}\gamma_\mu P_L b) (\bar{\ell} \gamma^\mu \gamma_5 \ell) \,
\label{eq:wilson-coefficients}.
\end{align}
To arrive at eq.~\eqref{eq:Lagrangian} we used CKM unitarity and neglected $V_{us}^* V_{ub} = O(\lambda^4)$ compared to $V_{ts}^* V_{tb} = O(\lambda^2)$ with the Wolfenstein parameter $\lambda \sim 0.22$, which is appropriate for the purposes of the present article.

\subsection{Kinematics}
The fully differential $\bar{B} \to X_s \ell^+ \ell^-$ rate is defined by three kinematical invariants, such as the dilepton mass square $q^2 = (p_{\ell^+} + p_{\ell^-})^2$, the dilepton energy $v\cdot q$, where $v = p_B/M_B$ is the heavy meson velocity which satisfies $v^2=1$, and an angular variable $z = \cos \theta$, where $\theta$ is the angle between the positively charged lepton momentum and the B meson momentum in the dilepton center of momentum frame,
\begin{align}
z &= \frac{v\cdot (p_{\ell^-} -p_{\ell^+})}{\sqrt{(v\cdot q)^2 - q^2}} \label{eq:z}\, .
\end{align}
We also define
\begin{align}
s = \frac{q^2}{m_b^2}\, , \indent u = \frac{(m_bv-q)^2}{m_b^2} \label{eq:s}\, ,
\end{align}
which appear in the calculation of the partonic decay process $b \to s(g) \ell^+ \ell^-$, in particular in the combinations (using similar notation as~\cite{Capdevila:2021vkf})
\begin{align}
w = 1-s\, , \indent \lambda = (w+u)^2-4u \,, \indent \mathcal{I} = \frac{1}{\sqrt{\lambda}} \ln \frac{w+u + \sqrt{\lambda} }{w+u-\sqrt{\lambda}}\, . \label{eq:I}
\end{align}
\subsection{Angular decomposition}
The branching fraction of $\bar{B} \to X_s \ell^+ \ell^-$ is quadratic in the angular variable $z$ at leading order in QED and can be decomposed into three angular observables~\cite{Lee:2006gs},
\begin{align}
\frac{d^3\mathcal{B}}{ds\,du\,dz} &= \frac{3}{8}\left[ (1+z^2) \frac{d^2 \mathcal{H}_T}{ds \, du}+ 2z \frac{d^2 \mathcal{H}_A}{ds \, du} + 2(1-z^2) \frac{d^2 \mathcal{H}_L}{ds \, du} \right] + O(\alpha_e) \, .
\end{align}
For the definitions of the angular observables in the presence of QED corrections, see~\cite{Huber:2015sra}. In the following, we work at lowest order in QED. The double differential branching fraction is given by
\begin{align}
\frac{d^2 \mathcal{B}}{ds \, du} &= \frac{d^2 \mathcal{H}_L}{ds \, du} + \frac{d^2 \mathcal{H}_T}{ds \, du} \label{eq:br}\,.
\end{align}
The angular observables depend on the Wilson coefficients according to\footnote{Interference terms involving the operator $Q_{10}$ are designated by the superscript 0}
\begin{align}
\frac{d^2 \mathcal{H}_T}{ds\, du} &= 2\Gamma_0 (1-s)^2 s \left[ (|C_9^{\rm eff}|^2 + C_{10}^2)\,h_T^{99}(s,u) + \frac{4}{s^2}|C_7^{\rm eff}|^2 \,h_T^{77}(s,u) \right. \nonumber \\
&\indent \left.+ \frac{4}{s} {\rm Re}(C_7^{{\rm eff}*} C_9^{\rm eff}) \, h_T^{79}(s,u) \right] + \frac{d^2 \mathcal{H}_T^{\rm brems}}{ds \, du}\, , \label{eq:HT}\\
\frac{d^2 \mathcal{H}_A}{ds \, du} &= -4 \Gamma_0 (1-s)^2 s \left[ {\rm Re}(C_9^{\rm eff}) C_{10} \, h_A^{90}(s,u) + \frac{2}{s} {\rm Re}(C_7^{\rm eff}) C_{10} \nonumber \, h_A^{70}(s,u) \right] \\
&\indent + \frac{d^2 \mathcal{H}_A^{\rm brems}}{ds\, du} \, , \label{eq:HA}\\
\frac{d^2 \mathcal{H}_L}{ds \, du} &= \Gamma_0 (1-s)^2 \left[ (|C_9^{\rm eff}|^2 + C_{10}^2)\,h_L^{99}(s,u)+ 4|C_7^{\rm eff}|^2 \,h_L^{77}(s,u) \right. \nonumber \\
& \indent \left. + 4{\rm Re}(C_7^{{\rm eff}*} C_9^{\rm eff}) \, h_L^{79}(s,u)\right] + \frac{d^2 \mathcal{H}_L^{\rm brems}}{ds \, du} \, ,\label{eq:HL}
\end{align}
where 
\begin{align}
\Gamma_0 = \frac{G_F^2 m_b^5}{48\pi^3 \tau_B^{-1}} |V_{tb}^* V_{ts}|^2 \label{eq:gamma0} \, .
\end{align} 
At tree level, the form factors are given by $h_I^{ij}(s,u)=\delta(u)$, so the $q^2$ dependence of the contribution of each product of coefficients to each observable can be understood by setting the form factors to unity. The effective coefficients $C_{7,9}^{\rm eff}$ absorb into $C_{7,9}$ the matrix elements of other operators in the effective theory which are proportional to the tree level matrix element of $Q_{7,9}$,
\begin{align}
C_7^{\rm eff}(s) &= C_7^{(11)}(\mu_b) \, , \\
C_9^{\rm eff}(s) &= C_9^{(11)}(\mu_b) + \frac{4\pi}{\alpha_s(\mu_b)} C_9^{(01)} (\mu_b) + \left(\frac{4}{3} C_1^{(00)}+C_2^{(00)}\right)  f_2 (s) \, .
\end{align}
Here $\mu_b \sim m_b$ and $C_i^{(nm)}$ is the coefficient of the $\alpha_s^n \kappa^m$ term of the double expansion of $C_i$ in $\alpha_s$ and $\kappa = \alpha_e/\alpha_s$~\cite{Huber:2005ig}. Including the $\alpha_s$ corrections to the Wilson coefficients would modify the hadronic mass spectrum which starts at order $\alpha_s$ at a higher order $\alpha_s^2$. Such matrix elements in the case of $C_7$ are order $\alpha_s$ and not included.  Below the charm pair production threshold, the loop function is real and is given by
\begin{align}
f_2(s) = \frac{4}{9} \left[ \ln \frac{\mu_b^2}{m_c^2} + \frac{2}{3} + y_c - (2+ y_c)\sqrt{y_c-1} \arctan \frac{1}{\sqrt{y_c - 1 }}\right] \, ,
\label{eq:f2}
\end{align}
where $y_c = 4m_c^2/q^2$. Finally, the bremsstrahlung terms in eqs.~\eqref{eq:HT}~--~\eqref{eq:HL} refer to the matrix elements of other operators in the effective theory which are not proportional to the tree level matrix elements of $Q_{7,9,10}$. 

We also define the normalized forward backward asymmetry $\bar{A}_{\rm FB}$ and fraction of transverse polarization $F_T$ according to
\begin{align}
\bar{A}_{\rm FB} = \frac{3}{4} \frac{\mathcal{H}_A}{\mathcal{H}_T + \mathcal{H}_L} , \indent F_T = \frac{\mathcal{H}_T}{\mathcal{H}_T + \mathcal{H}_L} \label{eq:afb} \, . \indent 
\end{align}
Both fractions are suppressed in the low-$q^2$ region due to the prefactor $s$ in eq.~(\ref{eq:HT}) for $\mathcal{H}_T$ and eq.~(\ref{eq:HA}) for $\mathcal{H}_A$, which does not appear in eq.~\eqref{eq:HL} for $\mathcal{H}_L$. Moreover, the two Wilson coefficients in the combination $C_9+2C_7/s$ appearing in eq.~\eqref{eq:HA} approximately cancel in the low-$q^2$ region, suppressing the forward backward asymmetry. The fraction $F_T$ is also suppressed since $\mathcal{H}_T$ depends on $C_9$ through the combination $|C_9+2C_7/s|^2$.
\subsection{Hadronic tensor and form factors}
\begin{figure}
\centering
\includegraphics[width=\textwidth]{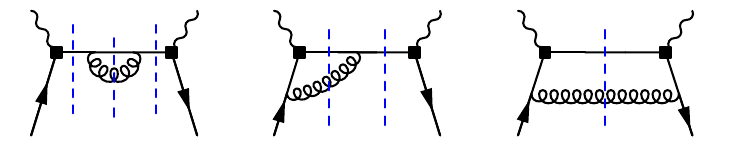}
\caption{The hadronic tensor at NLO from unitarity cuts. Square vertices are insertions of $Q_{7,9,10}$. Only one of the two vertex corrections is shown.
\label{fig:diagrams}}
\end{figure}

Having outlined the main ingredients in the phenomenology of $\bar{B} \to X_s \ell^+ \ell^-$, the starting point of a formal treatment of inclusive semileptonic decays in QCD is an analysis of the hadronic tensor, which is defined by the matrix element of the time ordered product of currents in QCD,
\begin{align}
W_{\mu \nu}^{ij}(v,q) = - \frac{1}{\pi}{\rm Im} \left[ -i \int d^4 x\, e^{-iqx}\frac{ \braket{\bar{B}(v)|TJ_\mu^{\dagger i}(x) J_\nu^j(0) |\bar{B}(v)}}{2M_B} \right].
\end{align}
The semileptonic and radiative currents are
\begin{align}
J_\mu^9 = J_\mu^{10} = \bar{s} \gamma_\mu P_L b \, , \indent J_\mu^7 = -\frac{2 m_b}{q^2} (\bar{s} i \sigma_{\mu \nu} P_R b)\, q^\nu \, .
\end{align}
Note that the hadronic currents for the operators $Q_9$ and $Q_{10}$ are identical. Therefore $W^{99}_{\mu \nu}=W^{90}_{\mu \nu}=W^{00}_{\mu\nu}$ and $W^{79}_{\mu \nu}=W^{70}_{\mu \nu}$. The hadronic tensor is a function of $v$ and $q$, and can be decomposed into five form factors
\begin{align}
W_{\mu \nu}^{ij}(v,q) &= -g_{\mu \nu} W_1^{ij}(q^2, v\cdot q) + v_\mu v_\nu W_2^{ij}(q^2, v\cdot q) + i \eps_{\mu \nu \alpha \beta}v^\alpha q^\beta W_3^{ij}(q^2, v\cdot q) \nonumber \\
&\indent + q_\mu q_\nu W_4^{ij}(q^2, v\cdot q) + (v_\mu q_\nu + v_\nu q_\mu)W_5^{ij}(q^2, v\cdot q)
\label{eq:decomp}
\end{align}
of which only the first three contribute to $\bar{B} \to X_s \ell^+ \ell^-$ for massless leptons. At next to leading order (NLO) the hadronic tensor is calculated from diagrams depicted in figure~\ref{fig:diagrams}. The form factors are related by
\begin{align}
&h_T^{99} = \frac{4\sqrt{\lambda}}{(1-s)^2} W_1^{99} \, , \indent h_T^{77} = \frac{\sqrt{\lambda} s^2}{(1-s)^2} W_1^{77}\, , \indent h_T^{79} = \frac{2\sqrt{\lambda}s}{(1-s)^2}W_1^{79} \, , \\[1em]
&h_A^{90} = \frac{2\lambda}{(1-s)^2}W_3^{99} \, , \indent h_A^{70} = \frac{\lambda s}{(1-s)^2} W_3^{79} \, , \\[1em]
&h_L^{99} = \frac{\sqrt{\lambda}(4s W_1^{99} + \lambda W_2^{99})}{(1-s)^2} \, , \,\,\,\, h_L^{77} = \frac{\sqrt{\lambda}(4sW_1^{77}+\lambda W_2^{77})}{4(1-s)^2} \, , \,\,\,\, h_L^{79} = \frac{\sqrt{\lambda}(4sW_1^{79} + \lambda W_2^{79})}{2(1-s)^2} \, .
\end{align}

\section{Results}
\label{sec:numerics}

In this section, we present results for the effect of a hadronic mass cut on the process $\bar{B} \to X_s \ell^+ \ell^-$ at low-$q^2$ at $O(\alpha_s)$. The SM predictions for $\bar{B} \to X_s \ell^+ \ell^-$ without a hadronic mass cut were updated recently in~\cite{Huber:2020vup}. For large values of $M_X^{\rm cut}$, the form factors can be expanded in local operators
\begin{align}
h_I^{ij} &= \sum_{n=0}^\infty \Big( \frac{\alpha_s}{4\pi} \Big)^n \left[ h_I^{ij(n)} + h_I^{ij (\lambda_1, n)} \frac{\lambda_1}{m_b^2} + h^{ij (\lambda_2, n)}_{I} \frac{\lambda_2}{m_b^2} \right] + O(1/m_b^3)  \, ,
\label{eq:ope}
\end{align}
where the normalization is such that, at lowest order, $h_I^{ij(0)} = \delta(u)$. The NLO corrections to the partonic decay rate are given by the functions $h_I^{ij(1)}$. Results for these functions are new, and have the form
\begin{align}
h_I^{ij(1)}(s,u) &= -4C_F \left[\frac{\ln u}{u} \right]_+ - C_F \Big( 7 - 8 \ln w \Big) \left[\frac{1}{u}\right]_+ + h_{I,\delta}^{ij(1)}\delta(u) + h_{I,\theta}^{ij(1)} \theta(u)
\label{eq:h1}
\end{align}
in terms of distributions in the partonic mass variable $u$. The coefficients of the plus distributions are universal, i.e.\ the same for each product of Wilson coefficients to each angular observable. The coefficients of the other distributions depend on the combination of Wilson coefficients which enter each angular observable. They are tabulated in Appendix~\ref{sec:matrix_elements}, and are also provided in electronic form in the supplementary materials.

One subtlety is that the expansion in terms of local operators in eq.~\eqref{eq:ope} requires $u\sim 1$. It is curious then, that these results are to be used for phenomenology of $\bar{B} \to X_s \ell^+ \ell^-$ with a cut $M_X < M_X^{\rm cut}$ which includes the region $u \ll 1$. However, the integral over the region $u\in[0,U]$ is to be interpreted as the difference of the bins $u \in [0, U_{\rm max}]$ and $u \in [U, U_{\rm \max}]$, where the first bin is the standard OPE integrated over the full hadronic mass spectrum, which admits a local expansion via analytic continuation to $u<0$, and the second is assumed to admit an OPE locally in $u\sim 1$. In practice, logarithms appear in the integrals
\begin{align}
\int_0^U du \left[ \frac{1}{u} \right]_+ = -\int_U^1 du \, \frac{1}{u} = \ln U \, , \indent \int_0^U du \left[ \frac{\ln u}{u} \right]_+ = -\int_U^1 du \, \frac{\ln u}{u} = \frac{1}{2}\ln ^2 U \, ,
\end{align}
using $[0,U] = [0,1]-[U,1]$ and the property that the integrals of the distributions vanish on the interval $[0,1]$, see Appendix~\ref{sec:plus}. If the hadronic mass cut is in the OPE region, i.e. $U \sim 1$, these logarithms are not large and do not require resummation. Therefore, each term in eq.~\eqref{eq:h1} is of equal importance.

The power corrections $h_I^{ij(\lambda_1,n)}$ are related to the leading power corrections $h_I^{ij(n)}$ order by order in perturbation theory through reparameterization invariance (RPI) relations~\cite{Manohar:2010sf}. We find, at lowest order,
\begin{align}
h_{I}^{ij(\lambda_1,0)} &= -\frac{1}{6}w^2\delta''(u) - \frac{1}{2}(2-w)\delta'(u) + h_{I,\delta}^{ij(\lambda_1,0)}\delta(u) \label{eq:hl0}\, 
\end{align}
and at NLO,
\begin{align}
&h_I^{ij(\lambda_1,1)} = \frac{4C_Fw^2}{3} \left[\frac{\ln u}{u^3} \right]_+ + \frac{C_Fw^2}{3}(1-8\ln w) \left[\frac{1}{u^3}\right]_+ \nonumber \\
&\indent  - 2C_F(2-w) \left[ \frac{\ln u}{u^2} \right]_+ - \frac{C_F}{6}\Big(2+3w - 24(2-w)\ln w \Big) \left[\frac{1}{u^2} \right]_+ \nonumber \\
&\indent + h_{I,[\ln/1]_+}^{ij(\lambda_1, 1)} \left[ \frac{\ln u}{u} \right]_+ \nonumber + h_{I,[/1]_+}^{ij(\lambda_1,1)} \left[ \frac{1}{u} \right]_+ \nonumber \\
 &\indent + h_{I,\delta\prime\prime}^{ij(\lambda_1,1)} \delta^{\prime\prime}(u) + h_{I,\delta\prime}^{ij(\lambda_1, 1)} \delta^\prime(u) + h_{I,\delta}^{ij(\lambda_1,1)} \delta(u) +h_{I,\theta}^{ij(\lambda_1,1)}\theta(u) \, .
\label{eq:hl1}
\end{align}
The first two lines of eq.~\eqref{eq:hl1} are generated exclusively from the derivatives of the plus distributions in eq.~\eqref{eq:h1}, and are therefore also universal. The second two lines contain pieces from the finite terms of eq.~\eqref{eq:h1}.

The calculation of the $h_{I}^{ij(\lambda_2,1)}$ corrections cannot be carried out using RPI relations, and is not attempted here. The structure of these corrections, however, can be inferred from the recent calculation of the hadronic mass distribution of $\bar{B} \to X_u \ell^- \nu$ at order $\alpha_s/m_b^2$~\cite{Capdevila:2021vkf}. The highest order plus distributions in $h_I^{ij(\lambda_2, 1)}$ are quadratic in $1/u$, rather than cubic as in the first line of eq.~\eqref{eq:hl1}. These quadratic divergences are not universal, as in the second line of eq.~\eqref{eq:hl1}. Schematically,
\begin{align}
h_I^{ij(\lambda_2,0)} = h_{I, \delta \prime}^{ij(\lambda_2,0)}\delta'(u) + h_{I, \delta}^{ij(\lambda_2,0)}\delta(u) \label{eq:hl2pre}
\end{align}
and
\begin{align}
&h_I^{ij(\lambda_2,1)} = h_{I,[\ln/2]_+}^{ij(\lambda_2,1)} \left[ \frac{\ln u}{u^2} \right]_+ + h_{I,[/2]_+}^{ij(\lambda_2, 1)} \left[\frac{1}{u^2} \right]_+ \nonumber \\
&\indent + h_{I,[\ln/1]_+}^{ij(\lambda_2,1)} \left[ \frac{\ln u}{u} \right]_+ \nonumber + h_{I,[/1]_+}^{ij(\lambda_2,1)} \left[ \frac{1}{u} \right]_+ \nonumber \\
 &\indent + h_{I,\delta \prime \prime}^{ij(\lambda_2,1)} \delta^{\prime\prime}(u) + h_{I,\delta\prime}^{ij(\lambda_2,1)} \delta^\prime(u) + h_{I,\delta}^{ij(\lambda_2,1)} \delta(u) +h_{I,\theta}^{ij(\lambda_2,1)}\theta(u) \, .
\label{eq:hl2}
\end{align}
To complete the calculation of the hadronic mass spectrum in the OPE at $O(\alpha_s/m_b^2)$, the coefficient functions in eq.~\eqref{eq:hl2} need to be computed in the future. The tree level functions in eq.~\eqref{eq:hl2pre} are available from~\cite{Ali:1996bm}.

\begin{figure}
\centering
\includegraphics[width=0.7\textwidth]{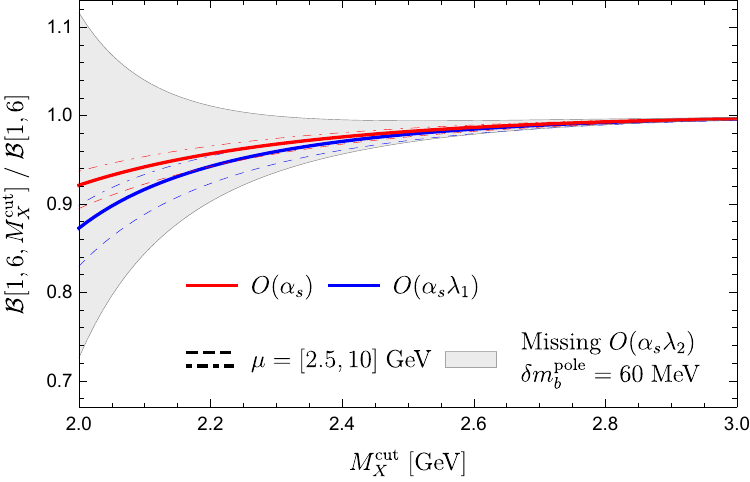}
\caption{Impact of a cut on the hadronic invariant mass on the low-$q^2$ branching fraction. Red and blue curves include up to $O(\alpha_s)$ and $O(\alpha_s \lambda_1/m_b^2)$ corrections, respectively. Dashed, solid and dot-dashed lines correspond to $\mu_b = [2.5,5,10]\; {\rm GeV}$, respectively. The gray band is a conservative bound on $O(\alpha_s \lambda_2/m_b^2)$ power corrections which are not yet available. \label{fig:BR} }
\end{figure}

The branching fraction of $\bar{B} \to X_s \ell^+ \ell^-$ integrated in a bin $q_1^2 < q^2 < q_2^2$, with and without a hadronic mass cut $M_X < M_X^{\rm cut}$, is
\begin{align}
\mathcal{B}[q_1^2, q_2^2, M_X^{\rm cut}] &= \int_{s_1}^{s_2} ds \int_0^{(1-\sqrt{s})^2} du \, \theta(M_X^{\rm cut}-M_X)\, \frac{d^2 \mathcal{B}}{ds\,du}  \, ,  \label{eq:bcut} \\
\mathcal{B}[q_1^2, q_2^2] &= \int_{s_1}^{s_2} ds \int_0^{(1-\sqrt{s})^2} du \, \frac{d^2 \mathcal{B}}{ds\,du} \label{eq:bnocut} \, , 
\end{align}
where $s,u$ are defined in eq.~\eqref{eq:s}, and 
\begin{align}
M_X = \sqrt{ u m_b^2 +m_b(M_B-m_b)(1-s+u) + (M_B-m_b)^2 }\label{eq:mx} \, .
\end{align}
The integral kernel in eqs.~\eqref{eq:bcut} and~\eqref{eq:bnocut} is given via eqs.~\eqref{eq:br}, \eqref{eq:HT} and~\eqref{eq:HL} in terms of form factors with an expansion in $\alpha_s$ and $1/m_b$ in eq.~\eqref{eq:ope}. The $O(\alpha_s)$, $O(\lambda_1/m_b^2)$ and $O(\alpha_s \lambda_1/m_b^2)$ terms of this expansion are given in eqs.~\eqref{eq:h1}, \eqref{eq:hl0} and~\eqref{eq:hl1}.

In the ratio $\mathcal{B}[q_1^2, q_2^2, M_X^{\rm cut}]/\mathcal{B}[q_1^2, q_2^2]$, the common factor in eq.~\eqref{eq:gamma0} cancels. Using the pole scheme for the heavy quark masses, the ratio depends on the parameters~\cite{ParticleDataGroup:2022pth}
\begin{align}
m_b = 4.78(6) \, {\rm GeV} \,, \,\,\,\, m_c = 1.67(7) \, {\rm GeV} \, , \,\,\,\, \alpha_s (M_Z) = 0.1179(9) \, , \,\,\,\,  M_B = 5.279\,{\rm GeV}
\end{align}
and the matrix element in the pole scheme~\cite{Gambino:2016jkc,Huber:2019iqf}
\begin{align}
\lambda_1 \equiv \frac{\braket{B|\bar{h}_v (iD_\perp)^2 h_v|B}}{2M_B} = -0.267(90)\,{\rm GeV}^2 ,
\end{align}
where $a_\perp^\mu = a^\mu - v^\mu (v\cdot a)$, 
in addition to the Wilson coefficients from~\cite{Huber:2005ig}. The pole scheme for heavy quark masses and the power correction parameter $\lambda_1$ is appropriate at this order since the conversions from a short distance scheme would affect the hadronic mass distribution which is order $\alpha_s$ at a higher order $\alpha_s^2$. The red curve in figure~\ref{fig:BR} shows the impact of a hadronic mass cut for $\mathcal{B}[1,6]$ at $O(\alpha_s)$. The blue curve includes also $O(\lambda_1/m_b^2)$ and $O(\alpha_s \lambda_1/m_b^2)$ corrections. The calculation is not complete at $O(\alpha_s/m_b^2)$ since $O(\alpha_s\lambda_2/m_b^2)$ corrections are not yet available. However, the missing $O(\alpha_s\lambda_2/m_b^2)$ corrections in eq.~\eqref{eq:hl2} do not contain $1/u^3$ plus distributions which appear for the $O(\alpha_s\lambda_1/m_b^2)$ ones in eq.~\eqref{eq:hl1}. Hence, at low $M_X^{\rm cut}$, the $O(\alpha_s \lambda_2/m_b^2)$ corrections are expected to be subdominant with respect to the $O(\alpha_s \lambda_1/m_b^2)$ ones. The grey band provides a conservative estimate of the missing power corrections at order $1/m_b^2$; it is centered on the $O(\alpha_s)$ result (red curved) and its width is twice the largest shift that we find by including $\lambda_1/m_b^2$ power corrections and varying $m_b^{\rm pole} = 4.78 (6) \; {\rm GeV}$. Although the tree level $\lambda_2$ corrections are available~\cite{Ali:1996bm}, they have not been included for simplicity since their effect is common to the branching fraction with and without a hadronic mass cut and cancel in their ratio $\mathcal{B}[q_1^2, q_2^2, M_X^{\rm cut}]/\mathcal{B}[q_1^2, q_2^2]$ at $O(1/m_b^2)$.

\begin{figure}
\centering
\includegraphics[width=0.49\textwidth]{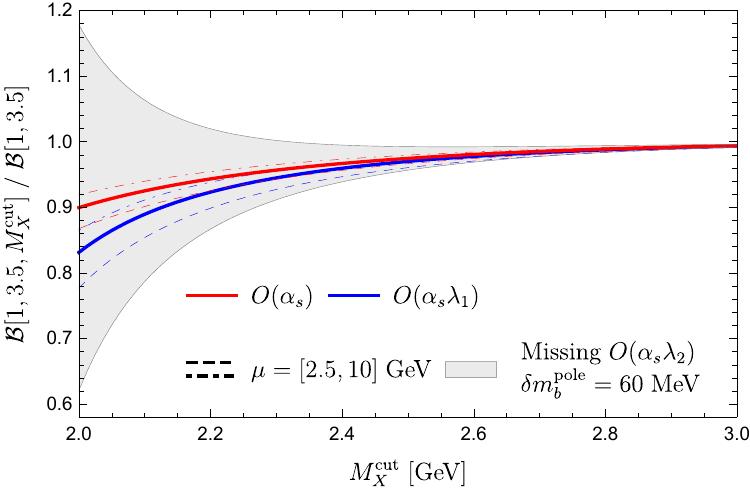}
\includegraphics[width=0.49\textwidth]{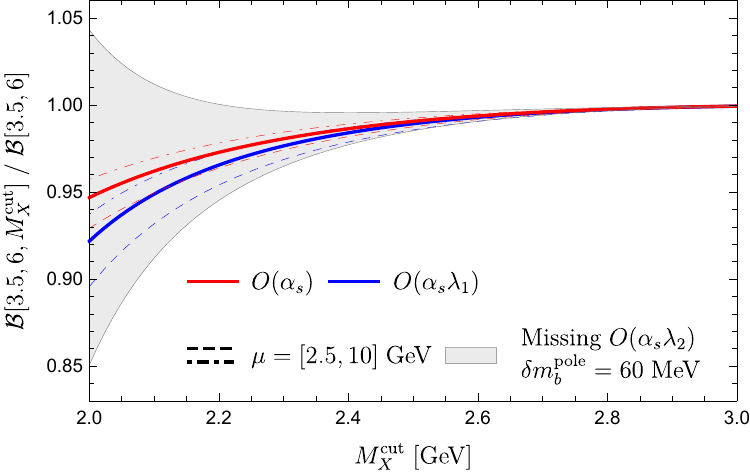}
\caption{Impact of a cut on the hadronic invariant mass on the low-$q^2$ branching fraction separated into two bins. See the caption in figure~\ref{fig:BR} for further details.
\label{fig:BRbins}}
\end{figure}

\begin{figure}
\centering
\includegraphics[width=0.49\textwidth]{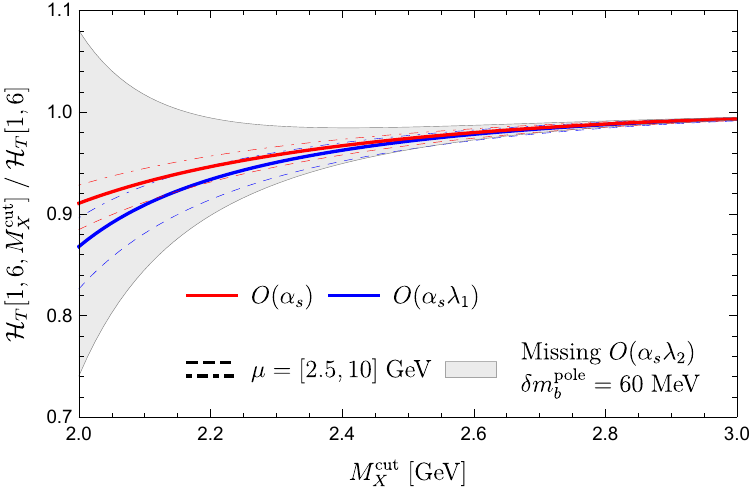}
\includegraphics[width=0.49\textwidth]{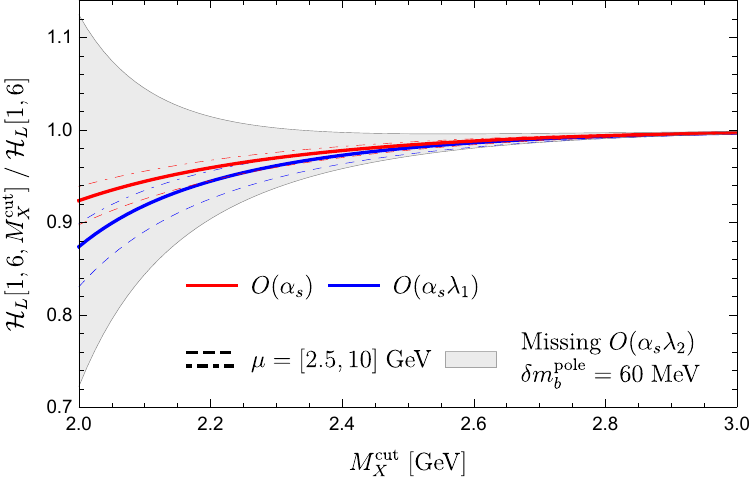}\\[0.6em]
\includegraphics[width=0.49\textwidth]{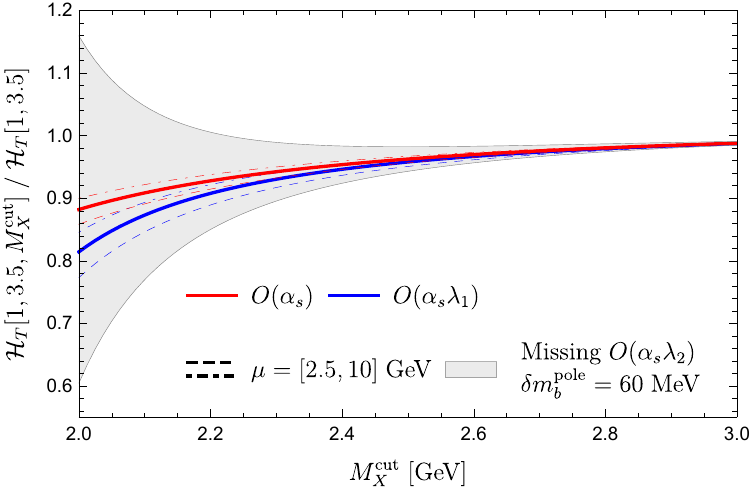}
\includegraphics[width=0.49\textwidth]{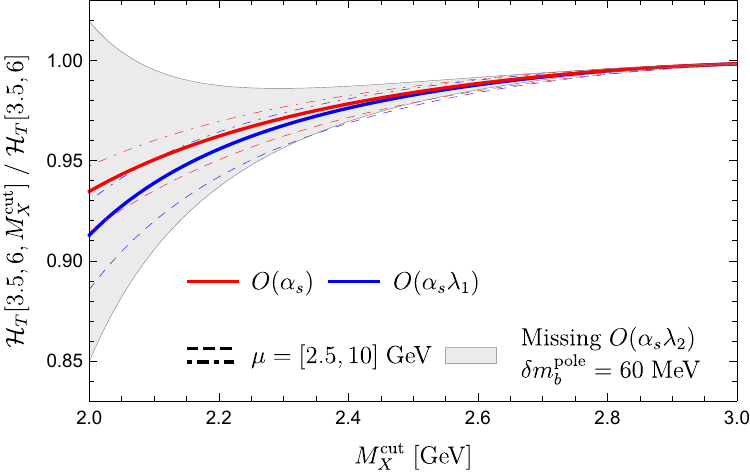}\\[0.6em]
\includegraphics[width=0.49\textwidth]{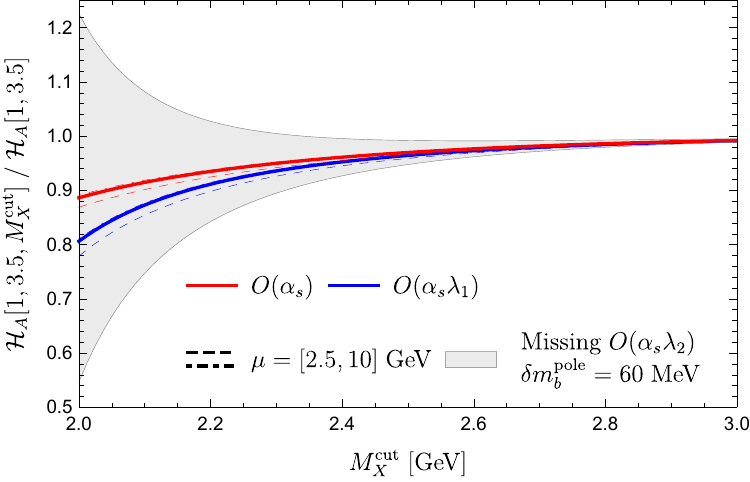}
\includegraphics[width=0.49\textwidth]{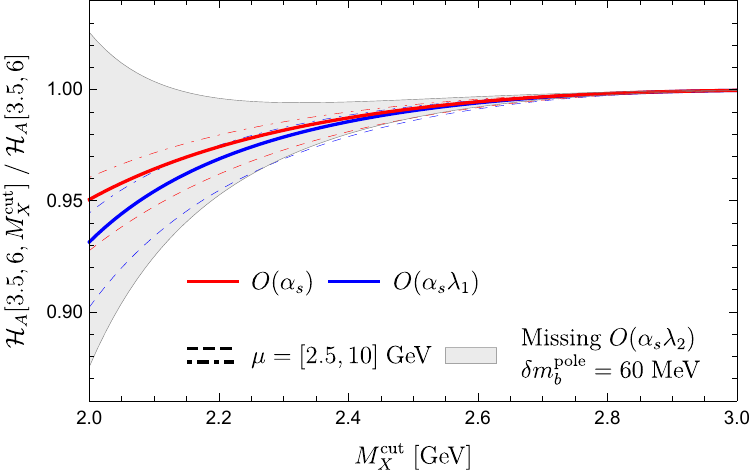}\\[0.6em]
\includegraphics[width=0.49\textwidth]{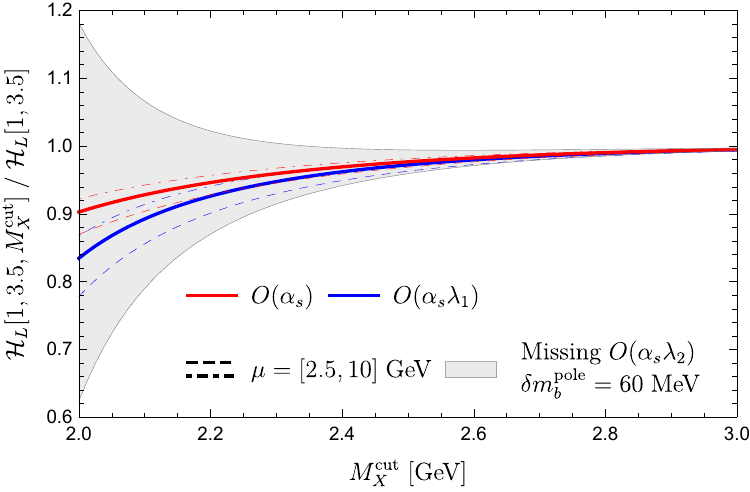}
\includegraphics[width=0.49\textwidth]{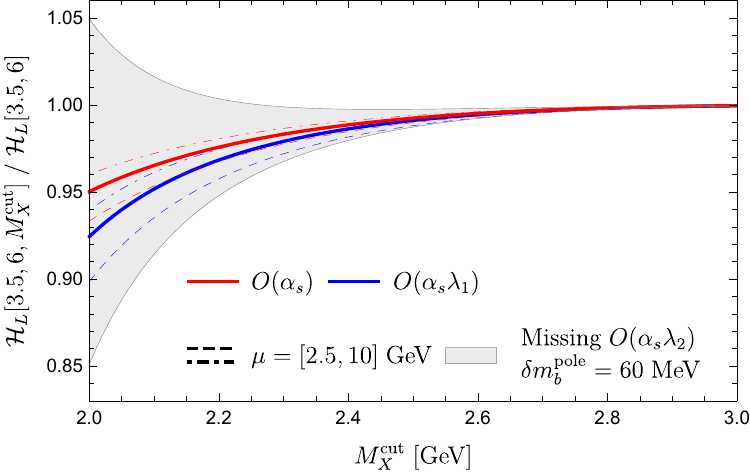}
\caption{Impact of a cut on the hadronic invariant mass on the low-$q^2$ angular observables. See the caption in figure~\ref{fig:BR} for further details. \label{fig:HI}}
\end{figure}

For $M_X^{\rm cut} = 2.0\,{\rm GeV}$, the $O(\alpha_s)$ correction decreases the integrated branching fraction by about 10\% and the $O(\alpha_s\lambda_1/m_b^2)$ effect is equally large and further lowers the branching fraction by another 10\%. It is important to stress that the large impact of the $\lambda_1$ correction at $M_X^{\rm cut} = 2.0 \, {\rm GeV}$ signals a breakdown of the OPE: the grey band cannot be interpreted as an estimate of power corrections to all orders in $1/m_b$ there, since their convergence is not guaranteed. A threshold  can be tentatively set at $M_X^{\rm cut} \gtrsim 2.5\; {\rm GeV}$, for which the grey band may be used to estimate the corrections from $\lambda_2$ as well as higher order power corrections. The same analysis for the branching fraction separated into two bins in the low-$q^2$ region is shown in figure~\ref{fig:BRbins}. The ratios $\mathcal{H}_I[q_1^2, q_2^2, M_X^{\rm cut}]/\mathcal{H}_I[q_1^2, q_2^2]$ for the angular observables can be built using eqs.~(\ref{eq:bcut}) and (\ref{eq:bnocut}) with $\mathcal{B} \to \mathcal{H}_I$, and are shown in figure~\ref{fig:HI}. The results presented in figures~\ref{fig:BR}, \ref{fig:BRbins} and \ref{fig:HI} are all very similar, and the discussion above for the branching fraction applies to these observables as well.

The qualitative difference between the cuts at $2.0\,{\rm GeV}$ and $2.5\,{\rm GeV}$ can be understood by converting from the hadronic mass to the partonic mass $\sqrt{u}m_b$, using eq.~\eqref{eq:mx}. The partonic mass corresponds to the offshellness of the strange quark in the OPE, and must be larger than the scale $\Lambda \sim 1\,{\rm GeV}$ in order for the OPE to make sense. For $M_X^{\rm cut} = 2.0\,{\rm GeV}$, putting $1\,{\rm GeV}^2 < q^2 < 6\,{\rm GeV}^2$, the partonic mass cut falls within the range $0.95\,{\rm GeV} < \sqrt{u_{\rm cut}}m_b < 1.21\,{\rm GeV}$. For $M_X^{\rm cut} = 2.5\,{\rm GeV}$, $1.70\,{\rm GeV} < \sqrt{u_{\rm cut}}m_b < 1.86\,{\rm GeV}$. The lower endpoint of this range is similar to the mass of the $\tau$ lepton, whose inclusive hadronic decays are analyzed with an OPE. Therefore, one may be optimistic that a cut $M_X^{\rm cut} = 2.5\, {\rm GeV}$ is sufficiently high to analyze $\bar{B} \to X_s \ell^+ \ell^-$ with an OPE as well. Our explicit results at $O(\alpha_s \lambda_1/m_b^2)$ presented in figures~\ref{fig:BR}, \ref{fig:BRbins} and \ref{fig:HI} support this conclusion.

\begin{figure}
\centering
\includegraphics[width=0.49\textwidth]{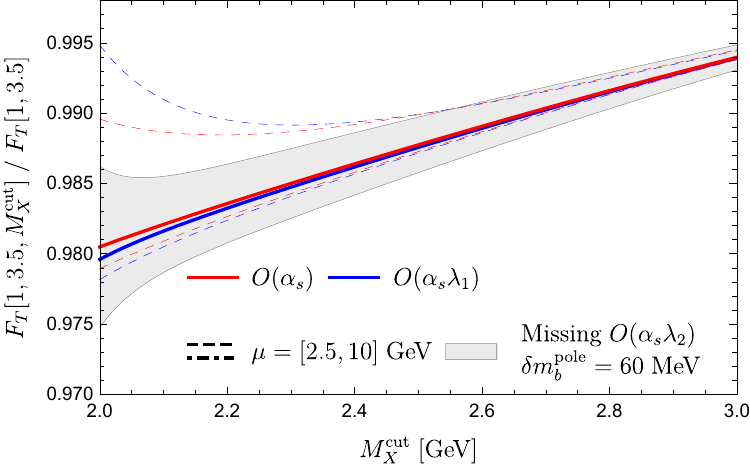}
\includegraphics[width=0.49\textwidth]{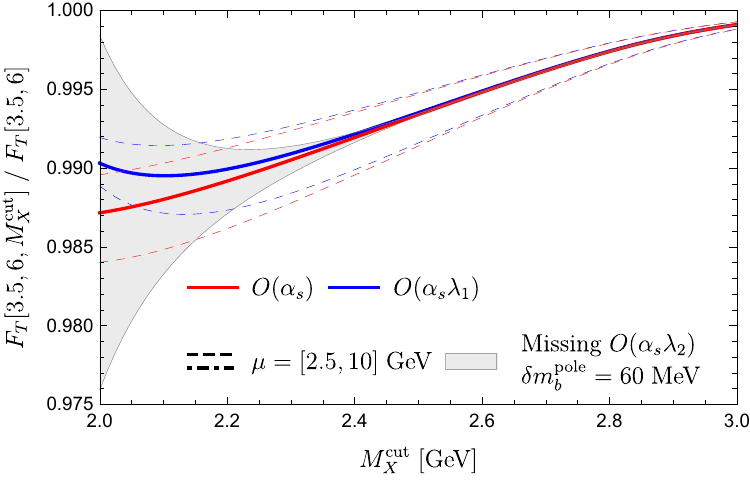}\\[0.6em]
\includegraphics[width=0.49\textwidth]{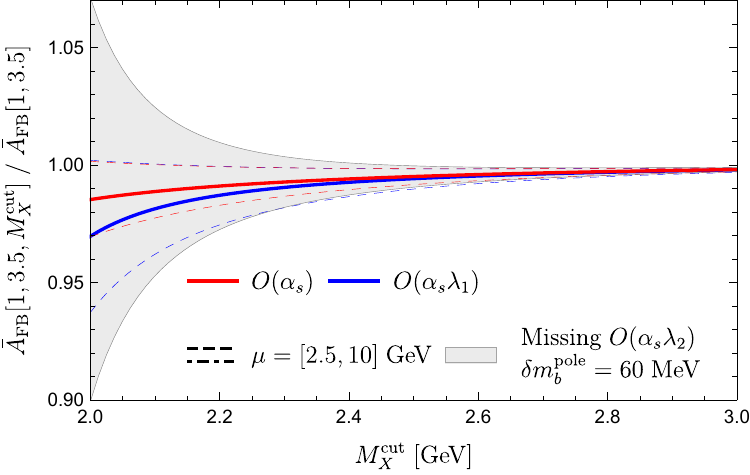}
\includegraphics[width=0.49\textwidth]{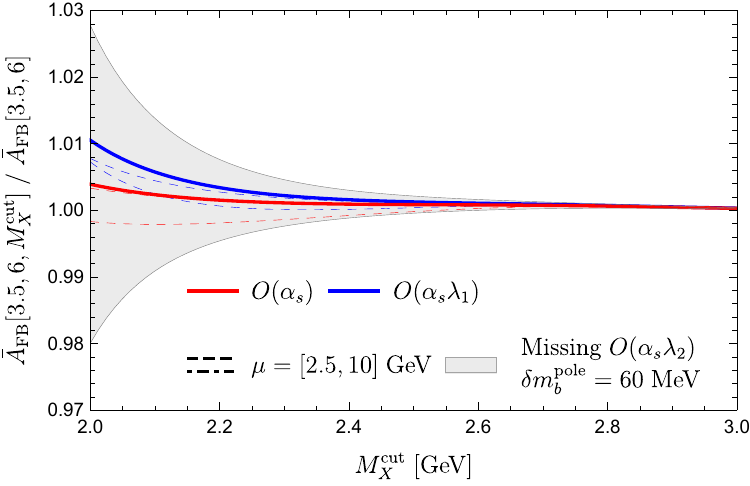}
\caption{Impact of a cut on the hadronic invariant mass on the low-$q^2$ normalized angular observables. See the caption in figure~\ref{fig:BR} for further details. \label{fig:FT_AFB}}
\end{figure}

In figure~\ref{fig:FT_AFB} we present our results for the cut dependence of the observables $F_T$ and $\bar A_{\rm FB}$. Due to limited statistics on $\bar{B} \to X_s \ell^+ \ell^-$, the numerators and denominators of eq.~\eqref{eq:afb} will probably need to be binned separately. To be precise,
\begin{align} 
\bar{A}_{\rm FB}[q_1^2, q_2^2, M_X^{\rm cut}] &= \ffrac{\int_{s_1}^{s_2} ds \int_0^{(1-\sqrt{s})^2} du \, \theta(M_X^{\rm cut}-M_X) \frac{3}{4} \frac{d^2 \mathcal{H}_A}{ds\,du} }{\int_{s_1}^{s_2} ds \int_0^{(1-\sqrt{s})^2} du  \, \theta(M_X^{\rm cut}-M_X) \frac{d^2\mathcal{B}}{ds\, du}} \, , \\
\bar{A}_{\rm FB}[q_1^2, q_2^2] &= \ffrac{\int_{s_1}^{s_2} ds \int_0^{(1-\sqrt{s})^2} du \, \frac{3}{4} \frac{d^2 \mathcal{H}_A}{ds\,du} }{\int_{s_1}^{s_2} ds \int_0^{(1-\sqrt{s})^2} du  \, \frac{d^2\mathcal{B}}{ds\, du}} \, ,
\end{align}
and
\begin{align}
F_T[q_1^2, q_2^2, M_X^{\rm cut}] &= \ffrac{\int_{s_1}^{s_2} ds \int_0^{(1-\sqrt{s})^2} du \, \theta(M_X^{\rm cut}-M_X) \frac{d^2 \mathcal{H}_T}{ds\,du} }{\int_{s_1}^{s_2} ds \int_0^{(1-\sqrt{s})^2} du  \, \theta(M_X^{\rm cut}-M_X) \frac{d^2\mathcal{B}}{ds\, du}} \, , \\
F_T[q_1^2, q_2^2] &= \ffrac{\int_{s_1}^{s_2} ds \int_0^{(1-\sqrt{s})^2} du \, \frac{d^2 \mathcal{H}_T}{ds\,du} }{\int_{s_1}^{s_2} ds \int_0^{(1-\sqrt{s})^2} du  \, \frac{d^2\mathcal{B}}{ds\, du}} \, .
\end{align}
The first important point to notice is that the $O(\alpha_s)$ corrections are $O(1\%)$, an order of magnitude smaller than for the rates presented in figures~\ref{fig:BR}, \ref{fig:BRbins} and \ref{fig:HI}. The reason for this behavior can be traced to the leading power results in eq.~(\ref{eq:h1}), where it is apparent that the $1/u$ singularities are universal to all angular observables. Upon integration, the resulting logarithms of $M_X^{\rm cut}$ appearing in the numerator and denominator of $F_T$ and $\bar A_{\rm FB}$ therefore cancel. Power corrections at $O(\alpha_s \lambda_1/m_b^2)$ are also quite small because $1/u^3$ and $1/u^2$ singular terms in eq.~\eqref{eq:hl1} are universal. This universality is lost for $O(\alpha_s \lambda_2/m_b^2)$ corrections. The latter, therefore, can potentially have a larger impact on these normalized ratios. In conclusion, the gray bands in figure~\ref{fig:FT_AFB} might underestimate the potential size of missing $O(\alpha_s \lambda_2/m_b^2)$ effects, especially at small values of $M_X^{\rm cut}$. This issue will be clarified when the $O(\alpha_s \lambda_2/m_b^2)$ corrections are available.

The panel for the observable $F_T[1,3.5]$ appears to be exceptional, in that even for $M_X^{\rm cut} = 2.0\,{\rm GeV}$, the $O(\alpha_s \lambda_1/m_b^2)$ correction is smaller than the $O(\alpha_s)$ correction, but this seems to originate from a cancellation that holds only at $\mu_b = 5\,{\rm GeV}$ (as it is apparent from inspection of the $\mu_b=2.5\; {\rm GeV}$ dashed curves). In general, the shifts observed as the scale $\mu_b$ is varied in the $[2.5,10]\; {\rm GeV}$ interval are also quite asymmetric for all observables because of various correlations between the numerator and denominators of $F_T$ and $\bar A_{\rm FB}$.

\begin{figure}
\centering
\includegraphics[width=0.7\textwidth]{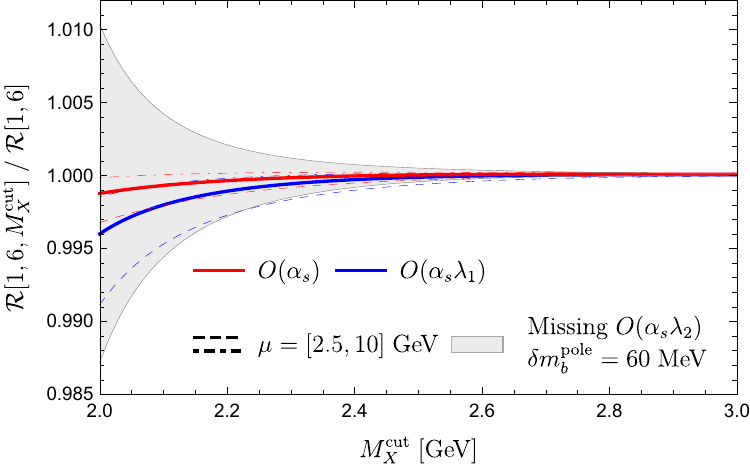}
\caption{Impact of a cut on the hadronic invariant mass on the low-$q^2$ observable $\mathcal{R}$. See the caption in figure~\ref{fig:BR} for further details. \label{fig:ZBR}}
\end{figure}

Finally, we consider the ratio formed by normalizing the ${\bar{B} \to X_s \ell^+ \ell^-}$ branching fraction to the branching fraction of $\bar{B} \to X_u \ell^- \nu$ with the same $q^2$ and $M_X$ cuts. Such a ratio was introduced in~\cite{Ligeti:2007sn} to analyze the high-$q^2$ region. Here we study its dependence on the hadronic mass cut in the low-$q^2$ region, as proposed in~\cite{Lee:2005pwa},
\begin{align}
\mathcal{R}[q_1^2, q_2^2, M_X^{\rm cut}] &= \ffrac{\int_{s_1}^{s_2} ds \int_0^{(1-\sqrt{s})^2} du \, \theta(M_X^{\rm cut} - M_X) \frac{d^2\mathcal{B}(\bar{B} \to X_s \ell^+ \ell^-)}{ds\,du}}{\int_{s_1}^{s_2} ds \int_0^{(1-\sqrt{s})^2} du  \, \theta(M_X^{\rm cut}-M_X) \frac{d^2\mathcal{B}(\bar{B} \to X_u \ell^- \nu)}{ds\,du}} \label{eq:zoltan} \, , \\
\mathcal{R}[q_1^2, q_2^2] &= \ffrac{\int_{s_1}^{s_2} ds \int_0^{(1-\sqrt{s})^2} du \, \frac{d^2\mathcal{B}(\bar{B} \to X_s \ell^+ \ell^-)}{ds\,du}}{\int_{s_1}^{s_2} ds \int_0^{(1-\sqrt{s})^2} du  \, \frac{d^2\mathcal{B}(\bar{B} \to X_u \ell^- \nu)}{ds\,du}} \, .
\end{align}
In the $C_7\to 0$ limit, $\bar{B}\to X_u \ell^-\nu$ and $\bar{B} \to X_s \ell^+\ell^-$ have the same hadronic mass spectra in QCD, to all orders in $\alpha_s$ and to all orders in $1/m_b$. While the $O(1\%)$ sensitivity of this observable to the hadronic mass cut shown in figure~\ref{fig:ZBR} is similar to that of $F_T$ and $\bar A_{\rm FB}$, it is important to observe that the cancellation of the $M_X^{\rm cut}$ dependence between numerators and denominators is different in the two cases. For the ratio $\mathcal{R}$, terms proportional to $|C_9^{\rm eff}|^2+|C_{10}|^2$ are completely independent of $M_X^{\rm cut}$ and the small cut dependence is controlled by terms involving $C_7$, which are subleading in the $\bar{B} \to X_s \ell^+ \ell^-$ rate by a factor $|4C_7C_9|/(C_9^2 + C_{10}^2) \sim 0.15$.

\section{Conclusions}
\label{sec:summary}

Measurements of the branching fractions and angular observables of inclusive $\bar{B} \to X_s \mu^+ \mu^-$ and $\bar{B} \to X_s e^+ e^-$, with precision competitive for the first time with SM predictions, are finally on the horizon~\cite{Belle-II:2018jsg}. To make the most of rare decay modes at Belle~II, it is important at this time to reflect on the results from the first generation B factories and pinpoint what observables are both straightforward to measure with high statistics and low backgrounds, and what observables are sensitive to physics beyond the SM with minimal interference from nonperturbative QCD. 

In this context, the most important conceptual advancement that can be made in the phenomenology of $\bar{B} \to X_s \ell^+ \ell^-$ at low-$q^2$ is to improve or replace the procedure of extrapolating out of the signal region $M_X \lsim 2.0 \,\rm{GeV}$, historically done on the experimental side, in order to compare with SM predictions without a hadronic mass cut. The observables $\bar{A}_{\rm FB}$, $F_T$ and $\mathcal{R}$ are significantly less sensitive to the hadronic mass cut than the standard angular observables, since they are formed from ratios of observables for which the hadronic mass spectrum is universal to an excellent approximation. The main conclusion of our study is that for $M_X^{\rm cut} \gsim 2.5\, {\rm GeV}$, this cut dependence may be calculated with an OPE. Since cuts in the OPE region are difficult to realize experimentally, a complementary calculation for cuts in the shape function region within SCET is planned, supplemented with an interpolation between shape function and OPE regions.

In conclusion, the angular observables $\mathcal{H}_T, \mathcal{H}_A, \mathcal{H}_L$ of $\bar{B} \to X_s \ell^+ \ell^-$ should be measured together with the branching fraction of $\bar{B} \to X_u \ell^- \nu$, with the same cuts on $q^2$ and $M_X$, without extrapolation in $M_X^{\rm cut}$. The data on $\bar{B} \to X_u \ell^- \nu$ is helpful in reducing the effect of the hadronic mass cut on the extraction of the Wilson coefficients of $\bar{B} \to X_s \ell^+ \ell^-$. A breakdown into bins of $M_X$ would be helpful, especially for the bins at $M_X \sim 2.0\,{\rm GeV}$ which are affected by large backgrounds. It is critical that all results are presented with correlations, since the ideal observables on the theory side are ratios of linear combinations of these branching fractions.


\subsubsection*{Acknowledgements}

The research of T.~Huber was supported by the Deutsche Forschungsgemeinschaft (DFG, German Research Foundation) under grant 396021762 -- TRR 257 ``Particle Physics Phenomenology after the Higgs Discovery.''
T.~Hurth is  supported by  the  Cluster  of  Excellence  ``Precision  Physics,  Fundamental
Interactions, and Structure of Matter" (PRISMA$^+$ EXC 2118/1) funded by the German Research Foundation (DFG) within the German Excellence Strategy (Project ID 39083149). T.~Hurth also thanks the CERN theory group for its hospitality during his regular visits to CERN where part of the work was done.


\begin{appendix}

\section{Analytical results}
\label{sec:analyticresults}

\subsection{Integrals}
The NLO matrix elements were reduced using IBP techniques with FeynCalc~\cite{Shtabovenko:2020gxv} and FIRE~\cite{Smirnov:2008iw}. The following master integral appears in the computation of the NLO matrix element:
\begin{align}
M_{a_1a_2a_3} &= -\frac{1}{\pi} \, \text{Im}\left[\frac{1}{D_0} \int \frac{[dk]}{D_1^{a_1}D_2^{a_2}D_3^{a_3}} \right] (m_b^2)^{a_1+a_2+a_3-1+\epsilon}\, ,
\end{align}
where $D_0 = p^2$, $D_1 = k^2$, $D_2 = (k-p)^2$ and $D_3 = (k+q)^2-m_b^2$, $p = m_b v - q$, the analytic continuation is specified by $D_i \to D_i + i0^+$, and the integral measure is ${[dk] = d^{4-2\eps}k \, e^{\epsilon \gamma_E}/(i\pi^{2-\eps})}$. After partial fractions and IBP reduction, the integrals needed are 
\begin{align}
M_{001} &= \left(\frac{1}{\eps} + 1 \right)\delta(u) \, , \label{eq:M001} \\
M_{101} &= \left(\frac{1}{\eps} + \frac{w \ln w}{1-w} + 2 \right)\delta(u) \, , \label{eq:M101} \\
M_{110} &= \left(\frac{1}{\eps} + 2 \right)\delta(u) - \left[ \frac{1}{u} \right]_+ \, , \label{eq:M110} \\
M_{111} &= -\frac{1}{w}\left(\Li_2(1-w)+2 \ln^2 w + \frac{\pi^2}{3} \right) \delta(u) + \left[\frac{\mathcal{I}}{u}\right]_+ \, \label{eq:M111} ,
\end{align}
where $\mathcal{I}$ is given in eq.~\eqref{eq:I}. In terms of standard distributions,
\begin{align}
\left[\frac{\mathcal{I}}{u} \right]_+ = - \frac{1}{w} \left[ \frac{\ln u}{u} \right]_+ + \frac{2 \ln w}{w} \left[ \frac{1}{u} \right]_+ + \frac{1}{u} \left( \mathcal{I} + \frac{1}{w} \ln \frac{u}{w^2} \right)\theta(u) \, .
\end{align}
All $1/\eps$ divergences are UV in origin. $M_{001}$ and $M_{101}$ are proportional to $\delta(u)$, since the loop integrals are real and elementary to evaluate. $M_{110}$ depends only on a single scale $u$ and is also elementary. For $u>0$, eq.~\eqref{eq:M111} can be confirmed by introducing Feynman parameters, using $-(1/\pi)\text{Im}[1/\Delta] = \delta(\Delta)$, and integrating over Feynman parameters, 
\begin{align}
M_{111}(u>0) &= \frac{1}{u} \int _0^1 dx_1 \int _0^1 dx_2 \,  \delta \Big[ x_1 \bar{x}_2 u - \bar{x}_1 x_2( \bar{x}_1 + x_1 w) \Big] = \frac{\mathcal{I}}{u}\, .
\end{align}
Here and in the following, for any variable $z$ we introduce $\bar{z} = 1-z$. The coefficients of the plus distributions are also inferred from this exercise by extracting the $u \to 0$ behavior of the $u>0$ result, obtaining an ansatz
\begin{align}
M_{111} &= M_{111}^\delta \delta(u) + \left[ \frac{\mathcal{I}}{u} \right]_+ \, \label{eq:ansatzM111}
\end{align}
in terms of an undetermined coefficient. To extract this coefficient, we combine the $1/u$ prefactor into the integral using a third Feynman parameter, integrate on the interval $u \in [0, U]$, remove the regulator $\eps$ ($U$ now serves as an IR regulator) and expand in $U \to 0$:
\begin{align}
&\int_0^U du\,M_{111}  = - \int_0^1 dx_1 \int _0^1 dx_2 \int _0^1 dx_3 \, \frac{\delta\Big[(x_1 \bar{x}_2 x_3 + \bar{x}_3) U -\bar{x}_1 x_2 x_3(\bar{x}_1+x_1 w)\Big]}{x_1 \bar{x}_2 x_3 + \bar{x}_3} \, \nonumber \\[0.3em]
& =  - \int_0^1 dx_1 \, \frac{1}{\bar{x}_1 (1-\bar{w} x_1)} \ln\Big(1+\frac{\bar{x}_1 (1-\bar{w} x_1)}{U \, x_1}\Big) \, \nonumber \\[0.3em]
& = - \! \MB{c_1}{z_1} \MB{c_2}{z_2} \, \frac{\Gamma(z_1+1)\Gamma(-z_2)\Gamma(1 + 2 z_1 - z_2)\Gamma^2(z_2-z_1)}{\Gamma(z_1+2)} \, U^{-z_1-1} \, w^{z_2}.
\end{align}
The Mellin-Barnes representation in the third line is well-suited for extracting the asymptotic behavior for $U \to 0$ to any desired order~\cite{Czakon:2005rk,MBasymptotics}. The values $c_1=-3/8$ and $c_2=-1/4$ denote the constant real parts of the integration contours in the complex plane. One obtains
\begin{align}
\int_0^U du\,M_{111} & =  -\frac{1}{w} \left( \Li_2(1-w) + 2\ln^2 w + \frac{\pi^2}{3} \right) - \frac{\ln^2 U}{2w} + \frac{2 \ln w}{w} \ln U + O(U) \, .
\end{align}
The constant term in this expression extracts the coefficient of the delta function in eq.~\eqref{eq:M111} respectively eq.~\eqref{eq:ansatzM111}.

\subsection{Form factors}
\label{sec:matrix_elements}
In this appendix, the eight form factors $h_{I}^{ij}$ are given as an expansion in $\alpha_s$ and $1/m_b$, up to $O(\alpha_s\lambda_1/m_b^2)$. The $O(\alpha_s)$ and $O(\alpha_s \lambda_1/m_b^2)$ corrections are proportional to $C_F = 4/3$. The following terms always appear together in the combination 
\begin{align}
L_w = \Li_2(1-w) + 2 \ln^2 w + \frac{\pi^2}{3} + \frac{1}{4} \Big\{ 0,1,2 \Big\} \ln \frac{\mu_b^2}{m_b^2} \, .
\end{align}
The bracket notation indicates the $ij=\{99, 79, 77 \}$ interference (in the case of $h_A^{ij}$, the last entry is deleted and $ij=\{90, 70\}$).
\subsubsection{$O(\lambda_1/m_b^2)$}

\begin{flalign}
& \Big\{ h_{T,\delta}^{99(\lambda_1,0)}, h_{T,\delta}^{79(\lambda_1,0)}, h_{T,\delta}^{77(\lambda_1,0)}\Big\} = \frac{1}{2} - \frac{4}{3w} \Big\{1, 0, w-1 \Big\} \, , &   \\
& \Big\{h_{A,\delta}^{90(\lambda_1,0)}\, , h_{A,\delta}^{70(\lambda_1,0)} \Big\} = \frac{1}{2} + \frac{4}{3w^2} \Big\{ 1-w, 1-w \Big\} \, , &  \\
&  \Big\{ h_{L,\delta}^{99(\lambda_1,0)}, h_{L,\delta}^{79(\lambda_1,0)}, h_{L,\delta}^{77(\lambda_1,0)}\Big\} = \frac{1}{2} - \frac{8}{3w} \Big\{w-1, 0 , 1 \Big\} \, .  &
\end{flalign}

\subsubsection{$O(\alpha_s)$}

\begin{flalign}
& \Big\{ h_{T,\delta}^{99(1)}, h_{T,\delta}^{79(1)}, h_{T,\delta}^{77(1)}\Big\} = -C_F \Big(4L_w - 8 \ln w  + 5  +  \Big\{2, 1, 0 \Big\} \frac{w \ln w}{1-w} \Big) \, , &  \\
& \Big\{ h_{A,\delta}^{90(1)}, h_{A,\delta}^{70(1)}\Big\} = -C_F \Big(4L_w - 8 \ln w  + 5 + \Big\{2, 1 \Big\} \frac{w \ln w}{1-w} \Big)  \, , \\
& \Big\{ h_{L,\delta}^{99(1)}, h_{L,\delta}^{79(1)}, h_{L,\delta}^{77(1)}\Big\} = -C_F \Big(4L_w - 8 \ln w  + 5 +  \Big\{0, 2, 4 \Big\} \frac{w \ln w}{1-w} \Big) \, , &  
\end{flalign}

\begin{flalign}
& \Big\{ h_{T,\theta}^{99(1)}, h_{T,\theta}^{79(1)}, h_{T,\theta}^{77(1)}\Big\} = \frac{4C_F}{u} \left[ \ln \frac{u}{w^2} + \sqrt{\lambda} \,\mathcal{I} + \frac{7}{4} \left(1-\frac{\sqrt{\lambda}}{w} \right) \right] & \nonumber \\
& \indent +\frac{2C_F\, \mathcal{I}}{w^2 \sqrt{\lambda}} 
\left\{
	\begin{array}{l} 
	2w^3 - w(14-5w)u - 4(2-w)u^2 + u^3 \, , \\
	2w^3 - w(9-4w)u + (1+2w)u^2 \, , \\
	2w^3 - w(4+w)u + 2(3-w)u^2 - u^3
	\end{array}
\right\} \nonumber \\
& \indent +\frac{C_F}{w^2 \sqrt{\lambda}} 
\left\{
	\begin{array}{l} 
	3w^2+4(3+w)u +u^2 \, , \\
	2(2-w)u - 2u^2 \, , \\
	-w^2(3-2w) - 4(1-w-w^2)u - (9-4w)u^2 + 2u^3
	\end{array}
\right\}  \, , \\
\nonumber \\
& \Big\{ h_{A,\theta}^{90(1)}, h_{A,\theta}^{70(1)} \Big\} = \frac{4C_F}{u} \left[ \ln \frac{u}{w^2} + w \,\mathcal{I} \right] - \frac{2C_F \mathcal{I}}{w^2}
\left\{
	\begin{array}{l}
	4w(1-w) + 3(2-w)u - u^2 \, ,\\
	4w(1-w) + (5-2w)u
	\end{array}
\right\} \nonumber \\
&\indent + \frac{C_F}{w^2}
\left\{
\begin{array}{l}
12-4w-u \, ,  \\
12-5w+2u \\
\end{array}
\right\}  \, , \\
\nonumber \\
& \Big\{ h_{L,\theta}^{99(1)}, h_{L,\theta}^{79(1)}, h_{L,\theta}^{77(1)}\Big\} = \frac{4C_F}{u} \left[ \ln \frac{u}{w^2} + \sqrt{\lambda} \,\mathcal{I} + \frac{7}{4} \left(1-\frac{\sqrt{\lambda}}{w} \right) \right] & \nonumber \\
& \indent +\frac{2C_F\, \mathcal{I}}{w^2 \sqrt{\lambda}} 
\left\{
	\begin{array}{l} 
	2w^3 + w(4-7w)u + 4(1-w)u^2 + u^3 \, , \\
	2w^3 - 3w(2-w)u + 2u^2 - u^3 \, , \\
	2w^3-w(16-5w)u-4(2-w)u^2+u^3   \\
	\end{array}
\right\} \nonumber \\
& \indent +\frac{C_F}{w^2 \sqrt{\lambda}} 
\left\{
	\begin{array}{l} 
	-w^2(7-4w) - 2(10-11w-2w^2)u - 3u^2 \, , \\
	-w^2 - 2(2-w)u + 3u^2 \, , \\
	5w^2+6(2+w)u+u^2   \\
	\end{array}
\right\} \, .
\end{flalign}

\subsubsection{$O(\alpha_s\lambda_1/m_b^2)$}

\begin{flalign}
& \Big\{ h_{T,[\ln/1]}^{99(\lambda_1, 1)} \, , h_{T,[\ln/1]}^{79(\lambda_1, 1)} \, , h_{T,[\ln/1]}^{77(\lambda_1, 1)} \Big\} = -2C_F \Big( 1- \frac{8}{3w}\Big\{1,0,w-1 \Big\} \Big) \, ,  &  \\
& \Big\{ h_{A,[\ln/1]}^{90(\lambda_1, 1)} \, , h_{A,[\ln/1]}^{70(\lambda_1, 1)} \Big\} = -2C_F \Big(1 + \frac{8}{3w^2} \Big\{1-w, 1-w \Big\} \Big)  \, ,  \\
& \Big\{ h_{L,[\ln/1]}^{99(\lambda_1, 1)} \, , h_{L,[\ln/1]}^{79(\lambda_1, 1)} \, , h_{L,[\ln/1]}^{77(\lambda_1, 1)} \Big\}  = -2C_F \Big( 1- \frac{16}{3w} \Big\{ w-1, 0, 1 \Big\} \Big)  \, , &
\end{flalign}

\begin{flalign}
& \Big\{ h_{T,[/1]}^{99(\lambda_1, 1)} \, , h_{T,[/1]}^{79(\lambda_1, 1)} \, , h_{T,[/1]}^{77(\lambda_1, 1)} \Big\} = \frac{C_F}{6w} \left\{
\begin{array}{l}
52-31w \, , \\
6 -33w \, , \\
-40+13w  
\end{array}
\right\} + \frac{4C_F}{3}
\left\{ 
\begin{array}{l}
-8+3w\, , \\ 
3w \, , \\
8-5w 
\end{array}
\right\} \frac{\ln w}{w} \, ,  &  \\
& \Big\{ h_{A,[/1]}^{90(\lambda_1, 1)} \, , h_{A,[/1]}^{70(\lambda_1, 1)} \Big\} = \frac{C_F}{6w^2} 
\left\{
\begin{array}{l}
-48+52w-31w^2 \, , \\
-48+54w-33w^2  
\end{array}
\right\}
+ \frac{4C_F}{3}
\left\{
\begin{array}{l}
8-8w+3w^2 \, ,  \\
8-8w+3w^2
\end{array}
\right\} \frac{\ln w}{w^2} \, ,  \\
& \Big\{ h_{L,[/1]}^{99(\lambda_1, 1)} \, , h_{L,[/1]}^{79(\lambda_1, 1)} \, , h_{L,[/1]}^{77(\lambda_1, 1)} \Big\} = \frac{C_F}{6w} \left\{
\begin{array}{l}
-80+57w \, , \\
12-35w \, , \\
104-31w  
\end{array}
\right\} + \frac{4C_F}{3}
\left\{
	\begin{array}{l} 
	16-13w \, , \\
	3w \, , \\
	-16+3w \,
	\end{array}
\right\} \frac{\ln w}{w} \, , &
\end{flalign}

\begin{flalign}
& \Big\{ h_{T,\delta\prime\prime}^{99(\lambda_1, 1)} \, , h_{T,\delta\prime\prime}^{79(\lambda_1, 1)} \, , h_{T,\delta\prime\prime}^{77(\lambda_1, 1)} \Big\} = \frac{2C_Fw^2}{3}\Big(L_w + \ln w - \frac{7}{8} + \frac{1}{4} \Big\{2, 1, 0 \Big\} \frac{w \ln w}{1-w} \Big)  \, , &  \\
& \Big\{ h_{A,\delta\prime\prime}^{90(\lambda_1, 1)} \, , h_{A,\delta\prime\prime}^{70(\lambda_1, 1)} \Big\} = \frac{2C_Fw^2}{3}\Big(L_w + \ln w - \frac{7}{8} + \frac{1}{4} \Big\{2, 1 \Big\} \frac{w \ln w}{1-w} \Big)  \, ,  \\
& \Big\{ h_{L,\delta\prime\prime}^{99(\lambda_1, 1)} \, , h_{L,\delta\prime\prime}^{79(\lambda_1, 1)} \, , h_{L,\delta\prime\prime}^{77(\lambda_1, 1)} \Big\}  = \frac{2C_Fw^2}{3}\Big(L_w + \ln w - \frac{7}{8} + \frac{1}{2} \Big\{0, 1, 2 \Big\} \frac{w \ln w}{1-w} \Big) \, ,  &
\end{flalign}

\begin{flalign}
& \Big\{ h_{T,\delta\prime}^{99(\lambda_1, 1)} \, , h_{T,\delta\prime}^{79(\lambda_1, 1)} \, , h_{T,\delta\prime}^{77(\lambda_1, 1)} \Big\}  = 2C_F(2-w) L_w  - \frac{C_F}{6} 
\left\{
	\begin{array}{l} 
	30-4w-10w^2 \, , \\
	30-7w - 10w^2 \, , \\
	30-10w-8w^2 \,   \\
	\end{array}
\right\} \nonumber \\
& \indent  + \frac{C_F}{6} \left\{
	\begin{array}{l} 
	32-44w+2w^2+16w^3 \, , \\
	32-50w+5w^2+16w^3 \, , \\
	32-56w+8w^2+16w^3 \,   \\
	\end{array}
\right\} \frac{\ln w}{1-w}  \, , & \\
& \Big\{ h_{A,\delta\prime}^{90(\lambda_1, 1)} \, , h_{A,\delta\prime}^{70(\lambda_1, 1)} \Big\}  = 2C_F(2-w) L_w  - \frac{C_F}{6} 
\left\{
	\begin{array}{l} 
	28-4w-10w^2 \, , \\
	28-5w - 10w^2 \\
	\end{array}
\right\} \nonumber \\
& \indent  + \frac{C_F}{6} \left\{
	\begin{array}{l} 
	32-44w+2w^2+16w^3 \, , \\
	32-50w+5w^2+16w^3 \\
	\end{array}
\right\} \frac{\ln w}{1-w}  \, , & \\
& \Big\{ h_{L,\delta\prime}^{99(\lambda_1, 1)} \, , h_{L,\delta\prime}^{79(\lambda_1, 1)} \, , h_{L,\delta\prime}^{77(\lambda_1, 1)} \Big\}  = 2C_F(2-w) L_w  - \frac{C_F}{6} 
\left\{
	\begin{array}{l} 
	30-14w-6w^2 \, , \\
	30-8w - 10w^2 \, , \\
	30-2w-10w^2 \,   \\
	\end{array}
\right\} \nonumber \\
& \indent  + \frac{C_F}{6} \left\{
	\begin{array}{l} 
	32-56w+8w^2+16w^3 \, , \\
	32-44w+2w^2+16w^3 \, , \\
	32-32w-4w^2+16w^3 \,   \\
	\end{array}
\right\} \frac{\ln w}{1-w}  \, , &
\end{flalign}

\begin{flalign}
& \Big\{ h_{T,\delta}^{99(\lambda_1, 1)} \, , h_{T,\delta}^{79(\lambda_1, 1)} \, , h_{T,\delta}^{77(\lambda_1, 1)} \Big\} = \frac{2C_F}{3w} L_w\Big\{8-3w, -3w, -8+5w \Big\} & \nonumber \\
& \indent  - \frac{C_F}{6w^2} \left\{
	\begin{array}{l} 
	14+2w-10w^2+9w^3+3w^4 \, , \\
	14+12w-8w^2+9w^3+3w^4 \, , \\
	14+22w-18w^2+9w^3+3w^4 \,   \\
	\end{array}
\right\} \nonumber \\
& \indent  - \frac{C_F}{6w} \left\{
	\begin{array}{l} 
	72-68w-22w^2+16w^3+8w^4 \, , \\
	4-4w-21w^2+16w^3+8w^4 \, , \\
	-64 +140w - 100w^2 + 16w^3+8w^4\,   \\
	\end{array}
\right\} \frac{\ln w}{1-w}  \, , \\
& \Big\{ h_{A,\delta}^{90(\lambda_1, 1)} \, , h_{A,\delta}^{70(\lambda_1, 1)} \Big\} = -\frac{2C_F}{3w^2} L_w (8-8w+3w^2) - \frac{C_F}{6w^2} \left\{
	\begin{array}{l} 
	28-12w^2+9w^3+3w^4 \, , \\
	28-4w-6w^2+9w^3+3w^4
	\end{array}
\right\} \nonumber \\
&\indent + \frac{C_F}{6w^2} \left\{
	\begin{array}{l} 
	48-120w+68w^2+22w^3-16w^4-8w^5 \, , \\
	48 - 116w + 68w^2 + 21w^3 - 16w^4 - 8w^5
	\end{array}
\right\} \frac{\ln w}{1-w}  \, , \\
& \Big\{ h_{L,\delta}^{99(\lambda_1, 1)} \, , h_{L,\delta}^{79(\lambda_1, 1)} \, , h_{L,\delta}^{77(\lambda_1, 1)} \Big\} = \frac{2C_F}{3w} L_w\Big\{-16+13w\, , -3w \, , 16-3w \Big\} & \nonumber \\
& \indent  - \frac{C_F}{6w^2} \left\{
	\begin{array}{l} 
	 14+14w-12w^2+5w^3+3w^4\, , \\
	 14-6w+9w^3+3w^4\, , \\
	 14-26w-12w^2+9w^3+3w^4\,   \\
	\end{array}
\right\} \nonumber \\
& \indent  - \frac{C_F}{6w} \left\{
	\begin{array}{l} 
	-96 + 212w - 140w^2 + 16w^3 + 8w^4\, , \\
	40-44w-14w^2+16w^3+8w^4 \, , \\
	176 - 172w - 16w^2 + 16w^3 + 8w^4\,   \\
	\end{array}
\right\} \frac{\ln w}{1-w} \, , 
\end{flalign}

\begin{flalign}
& \Big\{ h_{T,\theta}^{99(\lambda_1, 1)} \, , h_{T,\theta}^{79(\lambda_1, 1)} \, , h_{T,\theta}^{77(\lambda_1, 1)} \Big\} = -\frac{2C_F}{3wu^3}
\left\{
\begin{array}{l}
2w^3-3w(2-w)u + (8-3w)u^2 \, , \\
2w^3-3w(2-w)u - 3wu^2 \, , \\
2w^3 - 3w(2-w)u - (8-5w)u^2 \, 
\end{array}
\right\} \ln \frac{u}{w^2} & \nonumber \\
& \indent - \frac{C_F \mathcal{I}}{3w^2 u^3 \sqrt{\lambda}}
\left\{
\begin{array}{l}
4w^6 - 14w^4(2-w)u + 2w^2(24-16w+5w^2)u^2 \\
\,\, - w(36-36w+13w^2)u^3 - (8-58w+23w^2)u^4 \\
\,\, + 13(2-w)u^5 - 3u^6 \, , \\
4w^6 - 14w^4(2-w)u + 2w^2(24-24w+5w^2)u^2 \\
\,\, -w(6-17w+14w^2)u^3 - (4-41w+22w^2)u^4 \\
\,\, - 4(1+2w)u^5 \, , \\
4w^6 - 14w^4(2-w)u + 2w^2(24-32w+13w^2)u^2 \\
\,\, + w(24-34w+9w^2)u^3 + (16-8w+3w^2)u^4 \\
\,\, - (18-5w)u^5 + 3u^6
\end{array}
\right\} & \nonumber \\
& \indent -\frac{C_F}{6wu^3}
\left\{
\begin{array}{l}
2w^3-w(2+3w)u+(52-31w)u^2 \, , \\
2w^3-w(2+3w)u+(6-33w)u^2 \, , \\ 
2w^3-w(2+3w)u-(40-13w)u^2
\end{array}
\right\} \nonumber \\
& \indent + \frac{C_F}{6w^2 u^3 \sqrt{\lambda}} 
\left\{
\begin{array}{l}
2w^5-w^3(22-15w)u + 2w(28-14w-w^2)u^2 \\
\,\, -(64-110w+34w^2)u^3 +4(17-5w)u^4 - u^5 \, , \\
2w^5 - w^3(22-15w)u + 2w(28-37w-2w^2)u^2 \\
\, \, - (16-116w+43w^2)u^3 + 8(1-4w)u^4 - 6u^5 \, , \\
2w^5 - w^3(22-15w)u + 2w(28-60w+21w^2)u^2 \\
\,\, + 2(16+37w-22w^2+w^3)u^3 - 4(17-7w+w^2)u^4 \\
\, \, + (13-6w)u^5
\end{array}
\right\} \, ,  \\
& \Big\{ h_{A,\theta}^{90(\lambda_1, 1)} \, , h_{A,\theta}^{70(\lambda_1, 1)} \Big\} = -\frac{2C_F}{3w^2u^3}
\left\{
\begin{array}{l}
2w^4-3w^2(2-w)u - (8-8w+3w^2)u^2 \, , \\
2w^4-3w^2(2-w)u - (8-8w+3w^2)u^2
\end{array}
\right\} \ln \frac{u}{w^2} & \nonumber \\
& \indent - \frac{C_F \mathcal{I}}{3w^2 u^3}
\left\{
\begin{array}{l}
4w^5-10w^3(2-w)u-(12-24w+13w^2)u^3+10(2-w)u^4-3u^5 \, ,\\
4w^5-10w^3(2-w)u-(10-21w+14w^2)u^3 + 4(5-2w)u^4
\end{array}
\right\} & \nonumber \\
&\indent -\frac{C_F}{6w^2 u^2}
\left\{
\begin{array}{l}
16w^2(1-w) -8(1-w)(3-2w)u - (54-17w)u^2 + 7u^3 \, ,\\
16w^2(1-w) - 8(1-w)(3-2w)u - (52-22w)u^2 - 6u^3
\end{array}
\right\} \, ,  \\
& \Big\{ h_{L,\theta}^{99(\lambda_1, 1)} \, , h_{L,\theta}^{79(\lambda_1, 1)} \, , h_{L,\theta}^{77(\lambda_1, 1)} \Big\} = -\frac{2C_F}{3wu^3}
\left\{
\begin{array}{l}
2w^3-3w(2-w)u - (16-13w)u^2 \, , \\
2w^3-3w(2-w)u - 3wu^2 \, , \\
2w^3 - 3w(2-w)u + (16-3w)u^2 \, 
\end{array}
\right\} \ln \frac{u}{w^2} & \nonumber \\
& \indent - \frac{C_F \mathcal{I}}{3w^2 u^3 \sqrt{\lambda}}
\left\{
\begin{array}{l}
4w^6 - 14w^4(2-w)u + 2w^2(24-40w+21w^2)u^2 \\
\,\,+ 3w(24-38w+13w^2)u^3+(40-76w+45w^2)u^4 \\
\,\, - (22-19w)u^5 - 3u^6 \, ,\\
4w^6-14w^4(2-w)u + 2w^2(24-24w+5w^2)u^2 \\
\, \, + w(12+4w-15w^2)u^3 + w(22-21w)u^4 \\
\,\, - (2+3w)u^5 + 3u^6 \, ,\\
4w^6-14w^4(2-w)u+2w^2(24-8w+5w^2)u^2 \\
\,\, -w(48-58w+13w^2)u^3-(40-72w+23w^2)u^4 \\
\,\, +13(2-w)u^5 - 3u^6
\end{array}
\right\} &  \nonumber \\
& \indent - \frac{C_F}{6wu^3}
\left\{
\begin{array}{l}
2w^3 - w(2+3w)u - (80-57w)u^2  \, ,\\
2w^3 - w(2+3w)u + (12-35w)u^2  \, ,\\
2w^3 - w(2+3w)u + (104-31w)u^2
\end{array}
\right\}  \nonumber \\
& \indent + \frac{C_F}{6w^2 u^3 \sqrt{\lambda}}
\left\{
\begin{array}{l}
2w^5-w^3(22-15w)u + 2w(28-80w+43w^2)u^2 \\
\,\,+ 2(64-39w-15w^2+8w^3)u^3 \\
\,\, -(76-60w-16w^2)u^4 - 13u^5 \, ,\\
2w^5-w^3(22-15w)u +2w(28-34w-3w^2)u^2 \\
\, \, +(32+70w-52w^2)u^3 - 20(1+w)u^4 + 13u^5 \, ,\\
2w^5 - w^3(22-15w)u + 2w(28+12w-w^2)u^2 \\
\, \, - (64-90w+26w^2)u^3 + 12(3-w)u^4-u^5 
\end{array}
\right\} \, .
\end{flalign}

\subsection{Hadronic tensors}
\label{app:wi}

The leading order, $O(\alpha_s)$ and $O(\alpha_s \lambda_1/m_b^2)$ contributions to the functions $W_a^{ij}$ appearing in eq.~\eqref{eq:decomp} can be expressed as:
\begin{align}
W_a^{ij} &= W_a^{ij(0)} - \frac{\lambda_1}{2 m_b^2} W_a^{ij(\lambda_1,0)} + \frac{\alpha_s C_F}{4\pi} \left[ W_a^{ij(1)} - \frac{\lambda_1}{2 m_b^2} W_a^{ij(\lambda_1,1)}  \right] \; .
\end{align}
We present the explicit results for the leading order and $O(\alpha_s)$ contributions in terms of the variables $w= 1- \hat q^2$ and $u = (v - \hat q)^2$, where $\hat{q} = q/m_b$ and $v = p_B/M_B$. The tree level results are 
\begin{align}
W_a^{ij(0)} & = W_{a,\delta}^{ij(0)} \delta(u) \, ,
\end{align}
where
\begin{align}
\left\{ W_{1,\delta}^{99(0)}, W_{2,\delta}^{99(0)},W_{3,\delta}^{99(0)} \right\} &= 
\left\{ \frac{w}{4}, 1, \frac{1}{2} \right\} \, ,\\
\left\{ W_{1,\delta}^{79(0)}, W_{2,\delta}^{79(0)},W_{3,\delta}^{79(0)} \right\} &= \frac{1}{1-w}
\left\{\frac{w}{2}, 0, 1 \right\} \, ,\\
\left\{ W_{1,\delta}^{77(0)}, W_{2,\delta}^{77(0)},W_{3,\delta}^{77(0)} \right\} &= \frac{1}{(1-w)^2}
\left\{ w, -4(1-w), 2 \right\} \, .
\end{align}
The $O(\alpha_s)$ results are
\begin{align}
W_a^{ij(1)} = \; & W_{a,\delta}^{ij(0)} \; 
\left(  -4  \left[ \frac{\ln u}{u}\right]_+ + (8 \ln w -7) \left[ \frac{1}{u} \right]_+  + {\cal S}\; \delta(u) + \frac{4}{u} \ln \frac{u}{w^2} \; \theta(u) \right) \nonumber \\
 & + W_{a,\delta}^{ij(1)} \; \delta(u) + W_{a,\theta}^{ij(1)} \theta(u) \, ,
\label{eqapp:W}
\end{align}
where
\begin{align}
{\cal S}  = \; & -5 -\frac{4}{3}\pi^2 - 8 \ln^2 w-4 \Li_2 (1-w) \, .
\end{align}
The singular terms are
\begin{align}
\left\{ W_{1,\delta}^{99(1)}, W_{2,\delta}^{99(1)},W_{3,\delta}^{99(1)} \right\} =\; & 
 \frac{\ln w}{1-w} \left\{ \frac{w}{2}(4-5w), 10 (1-w), 4-5w \right\} \, , \\
\left\{ W_{1,\delta}^{79(1)}, W_{2,\delta}^{79(1)},W_{3,\delta}^{79(1)} \right\} =\; &
\frac{\ln w}{(1-w)^2} \left\{ \frac{w}{2} (8-9w) , -2 (1-w)  , 8-9w\right\} \nonumber \\
&\indent - \frac{1}{1-w} \left\{\frac{w}{2},0,1\right\}  \ln \frac{\mu_b^2}{m_b^2} \, , \\
\left\{ W_{1,\delta}^{77(1)}, W_{2,\delta}^{77(1)},W_{3,\delta}^{77(1)} \right\} =\; & 
 \frac{8\ln w}{(1-w)^2} \left\{ w,-6(1-w), 2 \right\} -\frac{2(2+w)}{(1-w)^2} \left\{0,0,1\right\} \, \nonumber \\
&\indent -  \frac{2}{(1-w)^2} \left\{w,-4(1-w),2\right\} \ln \frac{\mu_b^2}{m_b^2} \, .
\end{align}
The finite terms are
\begin{align}
W_{1,\theta}^{99(1)} =\; & 
\frac{{\cal I}}{2 u \lambda} (u^4+4 u^3 w-8 u^3+7 u^2 w^2-14 u^2 w+6 u w^3-8 u w^2+2 w^4) 
\nonumber \\ & 
+\frac{1}{4\lambda} (u^2+4 u w+12 u+3 w^2) \label{eq:W1t991} \, ,\\
W_{2,\theta}^{99(1)} =\; & 
\frac{2{\cal I}}{u \lambda^2}  (u^4 w+4 u^3 w^2-16 u^3 w+12 u^3+7 u^2 w^3-26 u^2 w^2+18 u^2 w+6 u w^4-8 u w^3
 \nonumber \\ & 
 +2 w^5)+ \frac{1}{\lambda^2} (7 u^3+29 u^2 w-60 u^2+43 u w^2-82 u w+80 u+21 w^3-38 w^2) \, , \\
W_{3,\theta}^{99(1)} =\; & 
\frac{{\cal I} }{u \lambda } (u^3+3 u^2 w-6 u^2+4 u w^2-4 u w+2 w^3)+\frac{1}{\lambda} (3 u+5 w-8) \, ,\\
W_{1,\theta}^{79(1)} =\; & 
\frac{{\cal I}}{u(1-w) \lambda} (2 u^3 w+u^3+6 u^2 w^2-9 u^2 w+6 u w^3-8 u w^2+2 w^4)
- \frac{u (u+w-2)}{ (1-w)\lambda} \, , \\
W_{2,\theta}^{79(1)} =\; & 
-\frac{4 u {\cal I}}{\lambda^2} (u^2+2 u w-u+w^2-3 w)
+  \frac{2}{\lambda^2}(5 u^2+4 u w-8 u-w^2) \label{eq:w2791} \, ,\\
W_{3,\theta}^{79(1)} =\; & 
\frac{2 {\cal I}}{u (1-w)\lambda } (2 u^2 w-5 u^2+4 u w^2-4 u w+2 w^3)
+\frac{9 u+9 w-16}{(1-w)\lambda } \, ,\\
W_{1,\theta}^{77(1)} =\; & 
-\frac{2{\cal I}}{u (1-w)^2 \lambda } (u^4+2 u^3 w-6 u^3-u^2 w^2+4 u^2 w-6 u w^3+8 u w^2-2 w^4)
\nonumber \\ & 
+\frac{1}{ (1-w)^2 \lambda} (2 u^3+4 u^2 w-9 u^2+4 u w^2+4 u w-4 u+2 w^3-3 w^2) \, ,\\
W_{2,\theta}^{77(1)} =\; & 
-\frac{8 {\cal I}}{u(1-w)\lambda^2} (u^4 w-2 u^4+4 u^3 w^2-14 u^3 w+14 u^3+7 u^2 w^3-22 u^2 
\nonumber \\ &
w^2+12 u^2 w+6 u w^4-8 u w^3+2 w^5)
- \frac{12}{(1-w)\lambda^2} (3 u^3+11 u^2 w-22 u^2
\nonumber \\ &
+15 u w^2-34 u w+32 u+7 w^3-12 w^2) \, ,\\
W_{3,\theta}^{77(1)} =\; & 
\frac{4 {\cal I}}{u(1-w)^2 \lambda} (u^3+u^2 w-4 u^2+4 u w^2-4 u w+2 w^3)
-\frac{4 (u-4) (u+w-2)}{ (1-w)^2 \lambda} \, . \label{eq:W3t771}
\end{align}
The results for $W_a^{99(1)}$ are identical to those presented in eqs.~(2.10) and (3.18)~--~(3.23) of~\cite{Capdevila:2021vkf}. Note that the $1/u$ singularity in the last term in the square bracket of eq.~(\ref{eqapp:W}) cancels against the singularities in $W_{a,\theta}^{ij(1)}$ (the last terms in the first bracket of eqs.~(\ref{eq:W1t991})~--~(\ref{eq:W3t771})); the absence of a singularity in eq.~\eqref{eq:w2791} reflects the absence of the corresponding singular term in eq.~(\ref{eqapp:W}) due to $W_{2,\delta}^{79(0)}=0$.

Explicit results for the $O(\alpha_s \lambda_1/m_b^2)$ terms $W_a^{ij(\lambda_1,1)}$ have been obtained using reparameterization invariance relations first presented in~\cite{Manohar:2010sf} and summarized in section~5 of~\cite{Capdevila:2021vkf}. As a cross check of these manipulations we verified that the terms $W_{a}^{99(\lambda_1,1)}$ reproduce exactly the results of~\cite{Capdevila:2021vkf} and that the integral of the $O(\alpha_s \lambda_1/m_b^2)$ corrections over the whole hadronic spectrum amounts to an overall factor $(1-\lambda_1/(2 m_b^2))$. 

\section{Plus distribution technology}
\label{sec:plus}
The plus distribution of order $(m, n)$, for integers $m\geq 0$ and $n \geq 1$ is defined by
\begin{align}
\int_{-\infty}^\infty du \left[ \frac{\ln^m u}{u^n} \right]_+ f(u) = \int_0^1 du\, \frac{\ln^mu}{u^n} \left(f(u) - \sum_{k=0}^{n-1} \frac{1}{k!}f^{(k)}(0) u^k\right)
\label{eq:plusdist}
\end{align}
for an arbitrary analytic function $f$. The simplification
\begin{align}
u^k \left[ \frac{\ln^ m u}{u^n} \right]_+ = \begin{cases}
\left[ \dfrac{\ln^m u}{u^{n-k}} \right]_+ & k<n
 \\[1em]
u^{k-n}\,\ln^m u \, \theta(u)\theta(1-u)& k \geq n
\end{cases}
\label{eq:pd2}
\end{align}
is consistent with the definition above, since multiplying both sides of eq.~\eqref{eq:pd2} by an analytic function and integrating over $\mathbb{R}$ gives equality by means of eq.~\eqref{eq:plusdist}. Generalizing,
\begin{align}
f(u) \left[ \frac{\ln ^m u}{u^n} \right]_+ &= \sum_{k=0}^{n-1} \frac{1}{k!}f^{(k)}(0) \left[ \frac{\ln^m u}{u^{n-k}} \right]_+ + \frac{\ln^m u}{u^n} \left(f(u) - \sum_{k=0}^{n-1} \frac{1}{k!}f^{(k)}(0) u^k \right) \theta(u)\theta(1-u) \label{eq:pd3} \, .
\end{align}

By definition, the $k^{\rm th}$ derivative $D^{(k)}$ of a distribution $D$ inherits the properties of $D$ after $k$ applications of integration by parts. Assuming no boundary conditions at infinity,
\begin{align}
\int_{-\infty}^{\infty} du\,D^{(k)}(u) f(u) &= (-1)^k \int_{-\infty}^{\infty} du\,D(u)f^{(k)} (u) \, .
\end{align}
To simplify expressions of the form $f(u) D^{(k)}(u)$, we integrate by parts
\begin{align}
\int_{-\infty}^{\infty} du \Big[ f(u) D^{(k)}(u) \Big] g(u) &= (-1)^k \int_{-\infty}^{\infty} du \, D(u) \frac{d^k}{du^k} \Big[f(u) g(u) \Big] \, ,
\end{align}
and then work (on a case by case basis) to remove the derivatives of $g(u)$ by integrating by parts again, so that $g(u)$ is restored and factors out of the integral. For instance,
\begin{align}
&\int_{-\infty}^{\infty} du \Big[ f(u)\,\delta'(u) \Big] g(u) = -\int_{-\infty}^{\infty} du\, \delta(u) \Big[ f(0)g'(u) + f'(0) g(u) \Big]\, \nonumber \\
&\indent = \int_{-\infty}^{\infty} du \Big[f(0) \delta'(u) - f'(0) \delta(u) \Big] g(u) \, ,
\end{align}
so $f(u)\delta'(u) = f(0) \delta'(u) - f'(0) \delta(u)$, which generalizes for higher derivatives to
\begin{align}
f(u)\delta^{(n)}(u) = \sum_{k=0}^n (-1)^k \binom{n}{k}f^{(k)}(0)\delta^{(n-k)}(u) \, .
\label{eq:pd4}
\end{align}
The derivative of a plus distribution is given in terms of higher order plus distributions and singular distributions,
\begin{align}
\frac{d}{du}\left[ \dfrac{\ln^m u}{u^n} \right]_+ &= \begin{cases}
- n \left[ \dfrac{1}{u^{n+1}} \right]_+ + \sum\limits_{k=0}^n \dfrac{(-1)^k}{k!} \delta^{(k)}(u) - \delta(1-u) & \indent m=0
\vspace{0.3cm}\\
 - n \left[ \dfrac{\ln^m u}{u^{n+1}} \right]_+ + m \left[ \dfrac{\ln^{m-1} u}{u^{n+1}} \right]_+ & \indent m>0 \, .\\
 \end{cases}
 \label{eq:pd5}
\end{align}
For instance,
\begin{align}
&\int_{-\infty}^{\infty} \frac{d}{du}\left[ \dfrac{1}{u} \right]_+ f(u) \, du = -\int_{-\infty}^{\infty} \left[ \dfrac{1}{u} \right]_+ f'(u)\, du  = -\int_0^1 \frac{f'(u) - f'(0)}{u} \, du \nonumber \\
&\indent = -\int_0^1 \frac{f(u) - f(0) - uf'(0)}{u^2} \, du - \left.\frac{f(u)-f(0)-uf'(0)}{u} \right|_0^1 \nonumber \\
&\indent = \int_{-\infty}^{\infty} \left( -\left[ \frac{1}{u^2} \right]_+ + \delta(u) - \delta'(u) - \delta(1-u)  \right)\,f(u)\,du \, .
\end{align} 
When the derivative of a plus distribution comes multiplied by an analytic function, the derivative is eliminated according to eq.~\eqref{eq:pd5}, and then eqs.~\eqref{eq:pd3} and~\eqref{eq:pd4} are used to further reduce the result to a standard form in which the coefficients of the plus distributions and Dirac distributions are independent of $u$. Moreover, second and higher order derivatives are obtained simply by iterating eq.~\eqref{eq:pd5}. Finally, derivatives of finite distributions generate singular distributions from their logarithmic singularities,
\begin{align}
\frac{d}{du}\Big[\ln^m u\, \theta(u) \theta(1-u) \Big] = m \left[\frac{\ln^{m-1}u}{u} \right]_+ \, .
\end{align}
For instance,
\begin{align}
&\int_{-\infty}^{\infty} \frac{d}{du} \Big[ \ln u \, \theta(u)\theta(1-u) \Big] f(u) \, du = - \int_{-\infty}^\infty \Big[ \ln u \, \theta(u)\theta(1-u) \Big] f'(u) \,du \nonumber \\
& \indent = -\int_0^1 \ln u \, f'(u) \, du = \int_0^1 \frac{f(u)-f(0)}{u}\, du + \Big.\ln u \, \Big(f(u)-f(0)\Big) \Big|_0^1 \nonumber \\
&\indent = \int_{-\infty}^{\infty} \left[ \frac{1}{u} \right]_+ f(u)\,du \, .
\end{align}

\end{appendix}


\bibliography{references}{}
\bibliographystyle{JHEP} 

\end{document}